\documentclass[%
 aip,
 amsmath,amssymb,
preprint,%
]{revtex4-2}
\usepackage{amsmath}
\DeclareMathOperator{\sech}{sech}

\DeclareMathOperator{\arccot}{arccot}

\DeclareMathOperator{\arccosh}{arccosh}

\usepackage{color}
\usepackage{appendix}
\usepackage{graphicx}% Include figure files
\usepackage{subfigure}
\usepackage{dcolumn}% Align table columns on decimal point
\usepackage{bm}% bold math
\usepackage[utf8]{inputenc}
\usepackage[T1]{fontenc}
\usepackage{mathptmx}
\usepackage{etoolbox}
\usepackage{ulem}

\makeatletter
\def\@email#1#2{%
 \endgroup
 \patchcmd{\titleblock@produce}
  {\frontmatter@RRAPformat}
  {\frontmatter@RRAPformat{\produce@RRAP{*#1\href{mailto:#2}{#2}}}\frontmatter@RRAPformat}
}%
\makeatother
\begin{document}

\preprint{AIP/123-QED}

\title[Frequency downshifting]{
Frequency downshifting in decaying wavetrains on the ocean surface covered by ice floes}
% Force line breaks with \\
\author{A.V.~Slunyaev}%
\affiliation{National Research University -- Higher School of Economics, 25 Bolshaya Pechorskaya Street, Nizhny Novgorod, 603950, Russia. %\\This line break forced with \textbackslash\textbackslash
}%
\affiliation{Institute of Applied Physics of the Russian Academy of Sciences, 46 Ulyanov Street, Box-120, Nizhny Novgorod, 603950, Russia.}%

\author{Y.A.~Stepanyants}
%\homepage{http://www.Second.institution.edu/~Charlie.Author.}
\affiliation{School of Mathematics, Physics, and Computing, University of Southern Queensland, 487--535 West St., Toowoomba, QLD 4350, Australia.}
\affiliation{Department of Applied Mathematics, Nizhny Novgorod State Technical University n.a. R.E. Alekseev, Nizhny Novgorod, 603950, Russia.}%
\email{Correspondin author: Yury.Stepanyants@unisq.edu.au}

\date{\today}% It is always \today, today, but any date may be explicitly specified

\begin{abstract}
We study analytically and numerically a frequency downshifting due to power-type frequency-dependent decay of surface waves in the ocean covered by ice floes. 
The downshifting is obtained both within the linear model and within the nonlinear Schr\"odinger (NLS) equation augmented by viscous terms for the initial condition in the form of an NLS envelope soliton. 
It is shown that the frequency-dependent dissipation produces a more substantial downshifting when the spectrum is relatively wide. 
As a result, the nonlinear adiabatic scenario of wavetrain evolution provides a downshifting remarkably smaller in magnitude than in the linear regime. 
Meanwhile, interactions between nonlinear wavegroups lead to spectral broadening and thus, result in fast substantial frequency downshifts. 
Analytic estimates are obtained for an arbitrary power $n$ of the dependence of a dissipation rate on frequency $\sim \omega^n$.
The developed theory is validated by the numerical modelling of the generalized NLS equation with dissipative terms. 
Estimates of frequency downshift are given for oceanic waves of realistic parameters.
\end{abstract}

\maketitle

\section{Introduction}
\label{Sect-1}

\begin{figure}[h!]
%\vspace{-0.5cm}
\centerline{\includegraphics[width=0.75\textwidth]{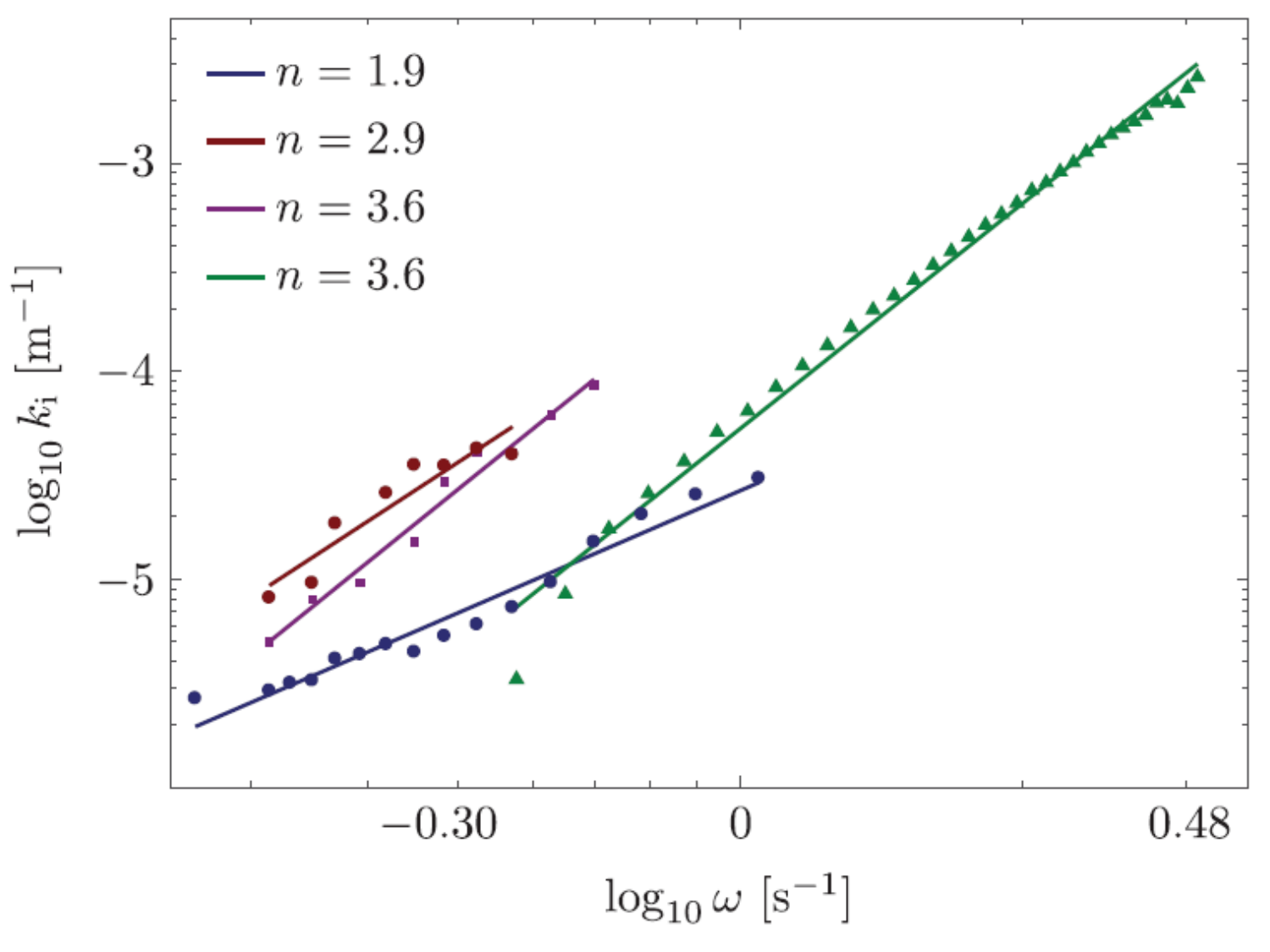}}%
\vspace{-0.5cm}
\caption{(Color online.) Median values of the imaginary part of the wavenumber as functions of frequency. 
Markers are data of measurements reported by different authors; solid lines are the best fits of corresponding data. Reproduced from Ref.~[\onlinecite{Meylan-18}].}
\label{Fig01}
\end{figure}

As is well-known, rather big parts of oceans in polar zones are covered by ice. 
By approaching polar regions, one can observe marginal ice zones (MIZ) covered by floes.
Surface oceanic waves entering MIZ experience a significant decay which essentially depends on the wave period. 
According to the experimental data, the dependence of the decay rate on the wave frequency has a power-type character $|k_i| \sim \omega^n$, where $k_i$ is the imaginary part of wavenumber and $\omega$ is the circular frequency, but there is a big uncertainty in the power $n$ reported by different authors as shown in Fig.~\ref{Fig01} from Ref.~[\onlinecite{Meylan-18}] (see also, Ref.~[\onlinecite{Hosekova-20}]).
Apparently, this uncertainty can be explained by the strong dependence of the decay rate on ice-floe parameters such as floe size, shape, thickness, and concentration on the ocean surface. 
Besides wave attenuation, a downshifting of spectrum maxima was observed in the course of wave propagation from the free ocean to the MIZ.
%****************************************************************
%****************************************************************

Understanding the changes that occur with ocean waves when they propagate in the ice-covered ocean is of significant importance and has applications to a range of problems, from marine safety and marine engineering to modeling ice extent. 
There are, however, many difficulties with measuring wave decay in the ice-covered ocean and interpreting the results. 
As aforementioned, the spatial decay rate of surface waves depends on many floe parameters. 
To obtain at least reasonable estimates, some authors take an average decay rate with the power three, $|k_i| \sim \omega^3$ (see, for example, Refs.~[\onlinecite{Alberello-22}], [\onlinecite{Alberello-23}]). 
Until now it is still unknown what is the mechanism by which oceanic waves lose energy (see, however, Refs.~[\onlinecite{Kheisin-1967}], [\onlinecite{Bukatov-2017}], [\onlinecite{Ni-2023}] where some models are suggested). 
This remains one of the biggest unsolved questions in geophysical fluid dynamics and, as expected, its importance grows with the lack of progress in its solution [\onlinecite{Kohout20}].

In relation to this, it is interesting to recall that the temporal decay rate of surface waves in clean deep water due to the influence of molecular viscosity is $\omega_i = -2\nu k^2$ (see, e.g., Refs.~[\onlinecite{Lamb-1932}], [\onlinecite{Landau-1987}]), where $\omega_i$ is the imaginary part of frequency, and $\nu$ is the water kinematic viscosity.
This can be presented in terms of the spatial decay rate $k_i = -4\nu\omega^5/g^3$, where $g = 9.81$ m/s$^2$ is the acceleration due to gravity.
It looks amassing that the dependence of the wave attenuation on frequency is stronger in clean water than in the case when the sea surface is covered by ice floes.

In the paper by Alberello \& P$\breve a$r$\breve a$u [\onlinecite{Alberello-22}], the authors presented results of a numerical study of a wave spectrum decay in the course of irregular oceanic waves propagation from clean water to the MIZ assuming that waves decay with the frequency-dependent spatial decay rate $k_i = -\mu\omega^3$.
They undertook numerical simulations for weakly nonlinear waves described by the nonlinear Schr\"odinger (NLS) equation augmented by a viscous term.
The calculations revealed the frequency downshifting but the amplitude dependence on distance was not exponential and rather irregular.
The irregularity was explained by the nonlinear interaction of wave components which can lead even to an increase in wave amplitude at certain distances.
This is true because, in the random wave field, envelope solitons can be formed and then, interact with each other resulting in occasional amplitude enhancement (see, for example, Ref.~[\onlinecite{Ducrozet-21}]).

In this paper, we reconsider the problem of frequency downshifting within the NLS equation assuming that the initial wavetrain has the shape of an envelope NLS soliton.
In such an idealized condition when the envelope soliton has been formed and can steadily travel in the absence of dissipation, we can focus on the ``pure effect'' of nonlinear frequency downshifting without the influence of other effects on this process.
Assuming that the dissipation is relatively weak in comparison to the nonlinear and dispersion effects, we develop an adiabatic theory to describe slowly decaying soliton in the course of its propagation. 
This approach allows us to reveal the frequency downshifting associated with the essentially nonlinear wave dynamics. We examine independently the linear and nonlinear regimes of wave attenuation accompanied by frequency shifting and estimate their strengths depending on real oceanic parameters.

The paper is organized as follows. The linear analytic description obtained for the general case and also under realistic assumptions is discussed in Sec.~\ref{Sect-2}, where the particular case of a sech-shape wavetrain is considered in more detail. 
The weakly nonlinear theory for modulated waves represented by the dissipative nonlinear Schr\"odinger (NLS) equation is introduced in Sec.~\ref{Sect-3}. The adiabatic theory for envelope solitons assuming that the effect of dissipation is weak is developed in Sec.~\ref{Sect-4}; it is compared with the results of the direct numerical simulation of the NLS equation in Sec.~\ref{Sect-5}. We further discuss and interpret the obtained results in Sec.~\ref{Sect-6} where the nonlinear mechanism of enhanced downshifting is suggested due to the interaction between solitary wavegroups. 
In the Appendix we discuss also the integrated amount of spectral energy flux. 
The main conclusions of the work are summarized in Sec.~\ref{Sect-7}.

\section{APPARENT FREQUENCY DOWNSHIFTING WITHIN THE FRAMEWORK OF THE LINEAR THEORY}
\label{Sect-2}

The frequency downshifting of modulated waves because of the frequency-dependent dissipation can occur within the framework of the linear theory.  
The general solution of the linear problem for the surface displacement $\eta(x,t)$ caused by waves traveling along the $x$-axis can be presented in the form:
\begin{eqnarray}
\eta(x,t) &=& \frac{1}{\sqrt{2\pi}} \int\limits_{-\infty}^{\infty}{\hat{\eta}(x,\omega) e^{i \omega t} d \omega}, \\
\hat{\eta}(x,\omega) &=&\hat{\eta}_0(\omega) e^{-ik x}, \quad
\hat{\eta}_0 (\omega) = \frac{1}{\sqrt{2\pi}} \int\limits_{-\infty}^{\infty}{{\eta}(0,t) e^{-i \omega t} d t}.    
\end{eqnarray}
Here $\eta_0(t)=\eta(0,t)$ is the boundary condition at some initial location $x=0$; $\hat{\eta}_0$ is its Fourier transform. The dispersion relation $k=k(\omega)$ is the necessary ingredient of the solution. 
It is convenient to use the quantity $S= 2|\hat{\eta}|^2$ defined for $\omega \ge 0$ which has the meaning of the frequency power spectrum of the wave field. 
When waves propagate in linear conservative media, this function remains unchanged, $S(\omega) = 2|\hat{\eta}(x,\omega)|^2=2|\hat{\eta}_0(\omega)|^2$.

The real part of the dispersion relation, $k_r(\omega)$, pertains to the case of waves in a perfect fluid, whereas in a real ocean waves experience attenuation in the course of propagation due to various dissipative factors.
Owing to this, the dispersion relation becomes complex, $k(\omega) = k_r(\omega) + ik_i(\omega)$, so that, the imaginary part of the wavenumber, $k_i(\omega)$, describes the spatial decay of a quasi-sinusoidal wave at a certain frequency $\omega$. Unfortunately, a theoretical derivation of a complex dispersion relation taking into account all possible dissipative factors in the ocean covered by floes is still challenging.
In such a situation, laboratory and field experiments are of high importance allowing us to determine the decay rate of surface waves as a function of wave frequency.
Surveys of experimental results on oceanic wave decay in the floe-covered surface can be found in Refs.~[\onlinecite{Meylan-18}], [\onlinecite{Squire-20}], [\onlinecite{Hosekova-20}] (see Fig.~\ref{Fig01}). 
Therefore, we have a complex dependence of the wavenumber on frequency, where $k_r(\omega) = \omega^2/g$ if the floe inertia is neglected (otherwise it is given by Eq.~(\ref{E02}) below), and $k_i(\omega)$ is empirically formulated as:
\begin{equation}
\label{E03}
k_i(\omega) = -\mu \omega^n, \quad \mbox{where} \; \mu = \frac{\rho_i}{\rho}\frac{h\nu}{g^2} \ge 0.
\end{equation}
Here $\rho_i$ and $\rho$ are the ice and water densities respectively, $h$ is the average floe thickness, and $\nu$ (in sec$^{n-4}$) is a coefficient that is determined by the ice floe concentration on the water surface [\onlinecite{Meylan-18}]; $n>0$ is the power of the frequency dependence of the dissipation.

The spectral function describing wave dissipation in the linear approximation is:
\begin{equation} 
\label{S}
S(x,\omega) = S_0(\omega) e^{-2\mu \omega^n x}, \quad \text {where} \quad S_0(\omega) = 2|\hat{\eta}_0(\omega)|^2.
\end{equation}

\subsection{The general theory of the linear transformation of a wave spectrum} \label{sec:GeneralLinearTheory}

Here we formulate a general analytic approach to study the spatial variation of a frequency spectrum, $S(\omega,x)$, of decaying waves with a specified dissipation law.
Let us first consider the evolution of a mean frequency $\omega_m$ defined as the ratio of two first spectral moments, $\omega_m = J_1/J_0$, where, in general, the spectral moments of the $n$-th order are defined as follows,
\begin{equation} 
\label{SpecMoments}
J_n = \int\limits_0^{\infty} S(x, \omega)\omega^n d\omega, \quad n = 0,\, 1,\, 2, \; \ldots \nonumber
\end{equation}
Then, the spatial rate of change of the mean frequency can be obtained by the direct calculation: 
\begin{equation}
\label{MeanFrequencyRate}
\frac{d}{dx} {\omega_m} = -2 \mu \left( \frac{J_{n+1}}{J_0} - \frac{J_1 J_n}{J_0^2} \right)    
\end{equation}
%\begin{eqnarray} 
%\label{MeanFrequencyRate}
%\frac{d}{dx} {\omega_m} &=& -2 \mu \left( \frac{m_{n+1}}{m_0} - \frac{m_1 m_n}{m_0^2} \right), \\
%\text{where} \quad {\omega_m} &=& \frac{m_1}{m_0} \quad \text{and} \quad m_l = \int\limits_0^{\infty} S(x, \omega)\omega^l d\omega, \quad l = 0,\, 1,\, 2, \; \ldots \nonumber
%\end{eqnarray}
%
The spectral moments for the given reference frequency $\omega_0$ can be rewritten using the binomial expansion of $\omega = \omega_0 \left(1 + \Delta \omega/\omega_0\right)$ where $\Delta \omega = \omega - \omega_0$:
\begin{align} 
\label{m_n}
J_n =  \omega_0^n \sum_{j\,=\,0}^{n}{\frac{n!}{j! (n-j)!} \int\limits_{0}^{\infty}{\left( \frac{\Delta\omega}{\omega_0} \right)^j}S(x,\omega)\, d\omega}.
\end{align}
If we now assume that the wave spectrum is narrow and concentrated in the vicinity of $\omega_0$, so that  
$$
\frac{|\Delta\omega|}{\omega_0} \ll 1 \quad \mbox{and} \quad
\frac{1}{J_0}\left|\int\limits_{0}^{\infty}{\left(\frac{\Delta \omega}{\omega_0} \right)^jS(x.\omega)\, d\omega}  \right| \sim \left| \frac{\Delta \omega}{\omega_0} \right|^j \ll 1, \quad \text{for}\, j=1,2,...,
$$
then the summands in Eq. (\ref{m_n}) are monotonically decaying when the number $j$ increases from $0$ to $n$. This observation helps to simplify the expression in brackets in Eq. (\ref{MeanFrequencyRate}). 
For the narrowband spectrum and choice of $\omega_0 = {\omega_m}$, Eq. (\ref{MeanFrequencyRate}) in the leading order of the relative spectral width $\delta_m$ reads:
\begin{align} 
\label{MeanFrequencyRate_NarrowBand}
\frac{1}{{\omega_m}} \frac{d}{d x} {\omega_m} = -2 n \mu \omega_m^n \left[\delta^2_{m}(x) + O(\delta_{m}^3) \right], \quad \mbox{where} \quad
\delta^2_{m}(x) = \frac{\int\limits_{0}^{\infty}{\left( {\omega-{\omega_m}} \right)^2}S(x,\omega)\, d\omega}{\omega_m^2 \int\limits_{0}^{\infty}{S(x,\omega)\, d\omega}}.
\end{align}

The width $\delta_{m}$ varies in the course of wave propagation, therefore Eq. 
(\ref{MeanFrequencyRate_NarrowBand}) for $\omega_m(x)$ with given $S_0(\omega)$ is integrodifferential and is hardly solvable in a closed form.
However, it provides an estimate of the frequency rate of change for any given initial spectrum $S_0(\omega)$.
In particular, it makes clear that the mean frequency decreases for any positive power $n$ of the frequency-dependent dissipation. 
This process occurs faster for a broader spectrum and a bigger value of $n$. 
Formula (\ref{MeanFrequencyRate_NarrowBand}) remains valid for non-integer powers $n$ too but in such a case, the binomial expansion (\ref{m_n}) ought to be replaced by the Taylor series, which has the same form as Eq. (\ref{m_n}), but contains formally an infinite number of summands.
	 
Alternatively, one can examine the evolution of the spectral peak frequency $\omega_p$, at which function $S(x,\omega)$ attains the maximum value for the given distance $x$. 
To this end, we differentiate function $S(x,\omega)$ in Eq. (\ref{S}) with respect to $\omega$ and equate the derivative to zero. 
The resultant equation is transcendental with respect to $\omega_p$ as a function of $x$. 
However, the inverse function can be readily obtained:
\begin{equation} \label{OmegaPeakCondition}
	x=\frac{1}{2 \mu n \omega_p^{n-1}} {\frac{d }{d \omega} \ln{S_0}} \Bigr|_{\omega=\omega_p}.
\end{equation}	 
Note that this formula does not require the knowledge of the variation of the spectral shape, thus it represents a self-consistent implicit solution for $\omega_p(x)$.

If the initial spectrum function $S_0(\omega)$ has the local maximum at $\omega = \omega_0$ (i.e., $\omega_0=\omega_p(0)$), then Eq. (\ref{OmegaPeakCondition}) yields the following approximate relation for the frequency $\omega_p$ provided that it does not differ much from $\omega_0$:
\begin{equation} \label{PeakFrequencyVariation_Initial}
\frac{1}{\omega_0} \frac{\omega_p(x)-\omega_0}{x} \approx \frac{1}{\omega_p} \frac{d \omega_p}{d x} \approx - 2n \mu \omega_p^{n} \delta_p^2, \quad \text{where} \quad \delta_p^2 = - \frac{S_0(\omega_0)}{\omega_0^2 S_0^{\prime \prime}(\omega_0)} .
\end{equation}%
This should occur, in particular, at the initial stage of the evolution of a wavetrain with the dominant frequency $\omega_0$. 
The estimate for the derivative of the peak frequency with respect to the distance (\ref{PeakFrequencyVariation_Initial}) coincides with the one for the mean frequency (\ref{MeanFrequencyRate_NarrowBand}) except for the definition of the spectral width $\delta_p$ instead of $\delta_m$. 
%Besides, the estimate (\ref{PeakFrequencyVariation_Initial}) does not use the assumption of the narrow-band spectrum. 
Besides, the assumption of the narrow-band spectrum is not used in the derivation of Eq. (\ref{PeakFrequencyVariation_Initial}). 
Both estimates predict that the frequency downshift is negative for $n>0$ and is proportional to the squared relative spectrum width. 
Formula (\ref{PeakFrequencyVariation_Initial}) can be also presented in the form:
\begin{equation} \label{PeakFrequencyVariation_Effective}
\left( \tilde{\omega_p} -1 \right) \approx  - 2n \xi \delta_p^2 \,,
\end{equation}
where $\tilde \omega = \omega/\omega_0$ and $\xi = \mu\omega_0^n x$. This is a linear approximation of the dependence of the frequency shift on the normalized distance $\xi$. 

In accordance with the attenuation of the spectral function (\ref{S}), one can write down for the peak frequency:
\begin{equation} \label{SpeactralDecay_Effective}
\frac{S(\omega_p(x))}{S_0(\omega_0)} = 	\frac{S_0(\omega_p(x))}{S_0(\omega_0)} e^{-2 {\tilde \omega}_p^n \xi} \,.
\end{equation}
Then, the solution (\ref{PeakFrequencyVariation_Effective}) can be used to estimate the decay of the maximum spectral amplitude for the given initial wave spectrum $S_0(\omega)$, which will be a function of the normalized coordinate $\xi$.

\subsection{An example of the linear transformation of a sech-shape wavetrain} 
\label{sec:ExampleLinearSech}

\begin{figure}[h]
%\vspace{-0.5cm}
%\centerline{\includegraphics[width=0.75\textwidth]{Fig02a.pdf}}%
\centerline{\includegraphics[width=1\textwidth]{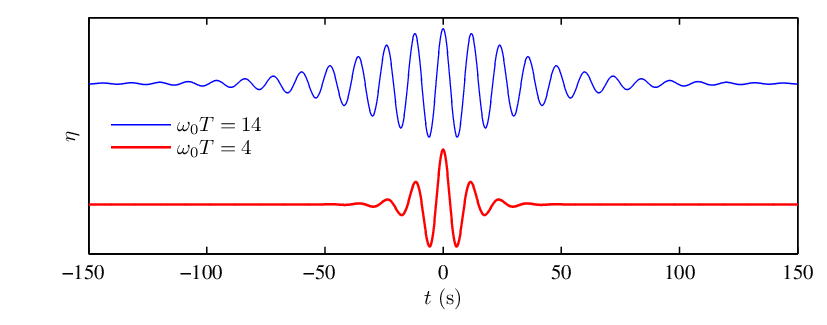}}%
\vspace{-0.5cm}
\caption{(Color online.) Wavetrains of the water surface perturbation as per Eq. (\ref{Eq24}) with the carrier wave period $12$~s and two characteristic durations, $T \approx 27$~s (above) and $T \approx 7.6$~s (below).}
\label{Fig02}
\end{figure}

To illustrate the concept of the frequency downshifting in the linear case, let us consider this effect in more detail for the initial condition taken in the form of a wavetrain with the {\it sech} shape: 
\begin{equation}%
\label{Eq24}%
\eta(x=0, t) =  A\sech{\left(t/T\right)}\cos{\omega_0 t} \equiv A\sech{\left(t/T\right)} \frac{e^{i \omega_0 t} + e^{-i \omega_0 t}}{2},
\end{equation}
where $\omega_0$ is the carrier wave frequency, $A$ is the wavetrain amplitude, $T$ is its characteristic time duration (see two examples in Fig. \ref{Fig02}). 
Such a shape of the wavetrain corresponds to the envelope soliton of the NLS equation that will be considered below.
%****************************************************************
%****************************************************************

\begin{figure}[h]
%\vspace{-0.5cm}
\centerline{\includegraphics[width=0.9\textwidth]{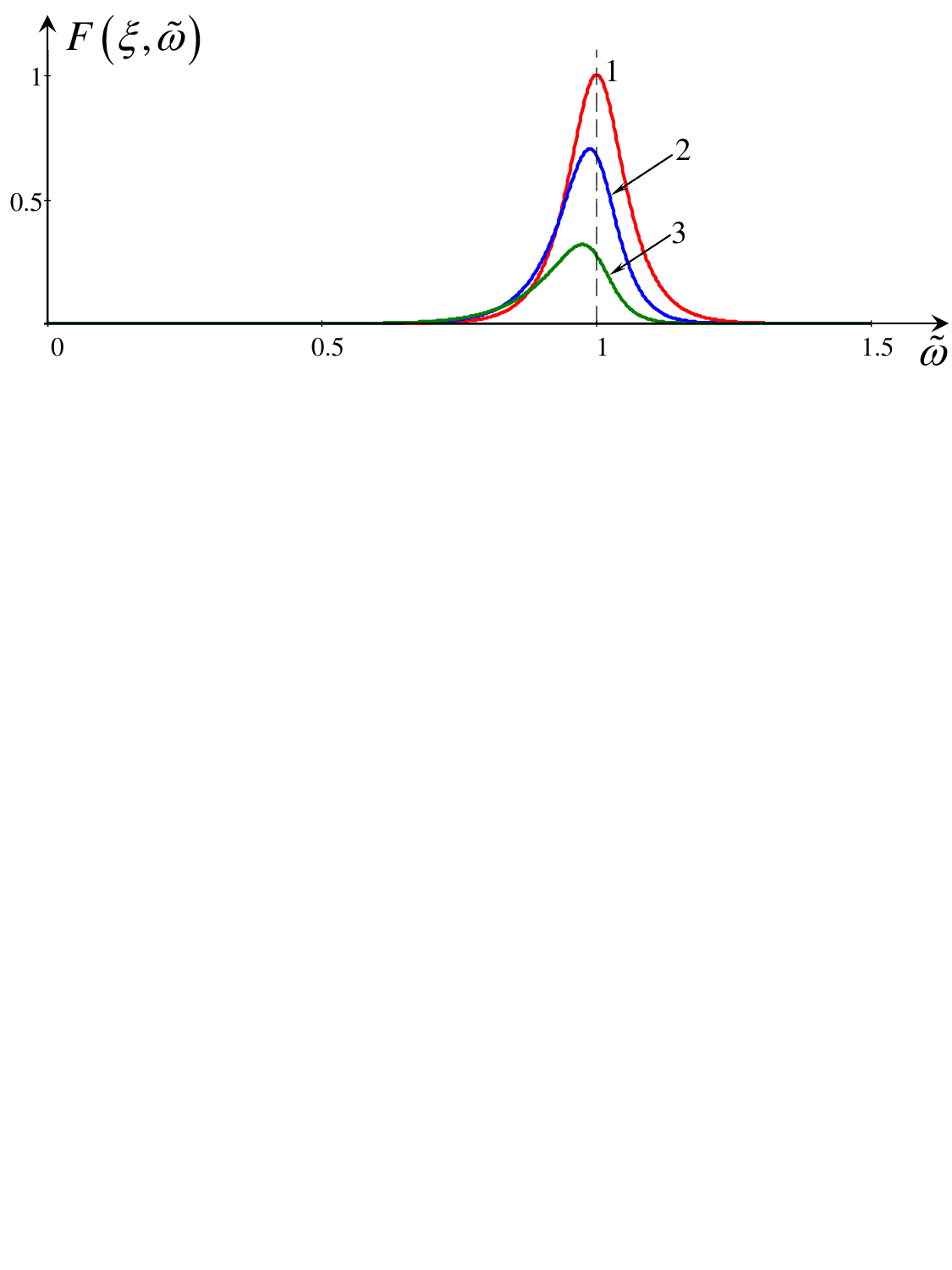}}%
\vspace{-14.0cm}
\caption{(Color online.) The normalized amplitude spectrum $F(\xi,\tilde{\omega}) = \left(2/AT\right) \sqrt{S(\xi,\tilde{\omega})/\pi}$ defined by Eq. (\ref{Eq26}) with $n = 3$ and $\omega_0 T \approx 14$ at different distances, $\xi = 0$ (line 1), $\xi = 2$ (line 2), and $\xi = 4$ (line 3). 
To make clearly visible the downshift of the spectra, the spectrum 2 was multiplied by the factor 5, and the spectrum 3 was multiplied by the factor 15.}
\label{Fig03}
\end{figure}

The Fourier transform of the wavetrain (\ref{Eq24}) is:
%
%\begin{widetext}
\begin{eqnarray}%
\hat{\eta}_0(\omega) &=&  \frac{AT}{2}\sqrt{\frac{\pi}{2}} \left[ \sech{\left( \left(\omega - \omega_0\right)\frac{\pi T}{2} \right) } + \sech{\left( \left(\omega + \omega_0\right)\frac{\pi T}{2} \right) }\right]. \label{Eq25}%
\end{eqnarray}
%\end{widetext}
%
The evolution of Fourier components due to the frequency-dependent wave attenuation can be written in the form
\begin{align}%
\label{Eq26}%
S(\xi,\tilde\omega) = \frac{\pi}{4}\left( AT \right)^2 \sech^2{\left[\left(\tilde \omega - 1\right)\frac{\pi \omega_0T}{2}\right]} \exp{\left(-2 {\tilde\omega}^n \xi\right)}.
\end{align}
This relation is strictly speaking not exact as the contribution from the second summand in Eq.~(\ref{Eq25}) is neglected for $\omega \ge 0$. 
Such a small inaccuracy can be disregarded unless the spectrum is unrealistically broad and slowly decaying.
The solution (\ref{Eq26}) is shown in Fig.~\ref{Fig03} for $n = 3$, $\omega_0 T \approx 14$, and three particular distances (cf. lines 2 and 3 with line 1). 
It is clear from the figure that in the course of the evolution the shape of the wave spectrum changes; it decays and becomes asymmetric. 
As a result, its maximum drifts to the left.
%****************************************************************
%****************************************************************

\begin{figure}[t!]
%\vspace{-0.5cm}
%\centerline{\includegraphics[width=1.0\textwidth]{Fig04.pdf}}%
%\vspace{-4.1cm}
\centerline{\includegraphics[width=0.5\textwidth]{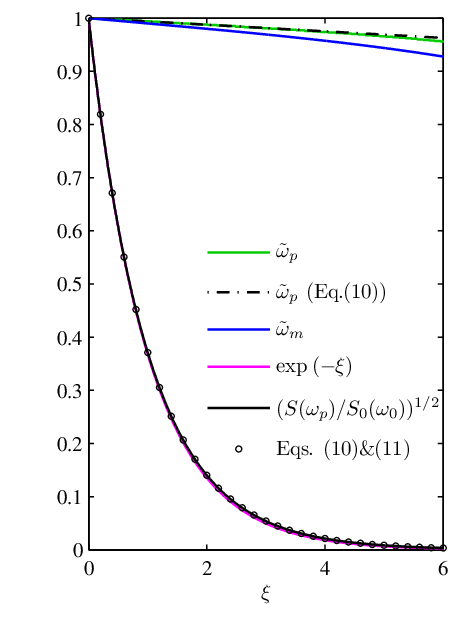} \includegraphics[width=0.5\textwidth]{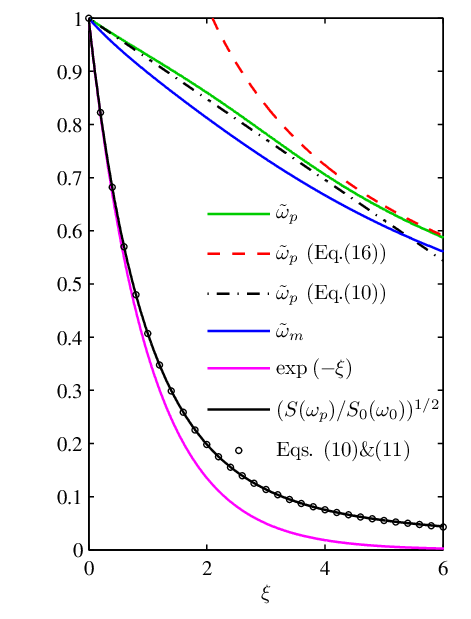}}%
\vspace{-0.5cm}
\phantom{WWWWWWW} (a) \phantom{WWWWWWWWWWWWWWWWWWWWW} (b) \phantom{WWWW}
\caption{(Color online.) Dependences of the dominant frequencies normalized by $\omega_0$ and the spectral amplitudes on the dimensionless distance within the linear theory for the case $n = 3$ and two characteristic widths of wavetrains shown in Fig.~\ref{Fig02}: $\omega_0 T=14$ (a) and $\omega_0 T=4$ (b).}
\label{Fig04}
\end{figure}

To examine the downshifting effect, we calculate the instantaneous frequency of the spectral peak 
$\tilde\omega_{p}$ where the maximum of the function $S(\xi, \tilde\omega)$ occurs at different values of normalized coordinates $\xi$, see Eq. (\ref{OmegaPeakCondition}). 
In this case, it is more convenient to differentiate the function $\sqrt{S(\xi, \tilde\omega)}$ with respect to $\tilde{\omega}$. The calculations eventually give the solution for $\xi$ as the function of the relative peak frequency $\tilde{\omega}_p$:
\begin{equation}%
\label{Eq27}%
\xi(\tilde\omega_{p}) = -\frac{\pi\omega_0 T}{2n\tilde\omega_{p}^{n-1}}\tanh{\left[\frac{\pi}{2}\omega_0T\left(\tilde \omega_{p} - 1\right)\right]}.
\end{equation}
The dependence (\ref{Eq27}) is shown in Fig.~\ref{Fig04} for $n = 3$ (see the green curves) for two values of the non-dimensional spectral width parameters $\omega_0 T$.
As shown in  Fig.~\ref{Fig02}, the choice $\omega_0 T =4$ corresponds to a very short wavetrain with a relatively broad spectrum; according to Fig.~\ref{Fig04}, the frequency shift is more significant in this case.
At the initial stage of the wave evolution, when $\tilde{\omega}_p \approx 1$, the downshifting can be estimated by expanding the function $\tanh{(\cdot)}$ in Eq. (\ref{Eq27}) or with the use of general relations  (\ref{MeanFrequencyRate_NarrowBand}) or (\ref{PeakFrequencyVariation_Initial}). 
The relative spectral width parameters $\delta_m$ and $\delta_p$ for the given wavetrain shape (\ref{Eq26}) are only slightly different: $\delta_m^2 = 1/[3(\omega_0 T)^{2}]$ and $\delta_p^2 = 2/(\pi \omega_0 T)^{2}$, i.e. $\delta_m/\delta_p = \pi/\sqrt{6} \approx 1.28$. 
As follows from Fig.~\ref{Fig04}, the dependencies for $\omega_p(\xi)$ and numerically calculated $\omega_m(\xi)$ generally agree; the latter follows slightly below, in accordance with the estimates of $\delta_m$ and $\delta_p$.
%
%****************************************************************
%****************************************************************
At a large distance when the peak frequency becomes much smaller than it was originally, $\tilde{\omega}_p \ll \omega_0$, the relation (\ref{Eq27}) simplifies and yields:
\begin{equation} \label{PeakFrequencyVariation_Asymptotics}
\tilde{\omega}_p \approx \left[\frac{\pi \omega_0T \tanh{\left(\pi \omega_0 T/2\right)}}{2n \xi} \right]^{\frac{1}{n-1}}.
\end{equation}
This dependence can be further simplified if $\omega_0 T \gtrsim 1$; then $\tanh{\left(\pi \omega_0 T/2\right)} \approx 1$, and we have
$\tilde{\omega}_p \approx \left(\pi \omega_0T /(2n \xi) \right)^{\frac{1}{n-1}}$. 
In particular, for $n=3$ the asymptotic dependence of the peak frequency on the distance, as per Eq. (\ref{PeakFrequencyVariation_Asymptotics}), is $\tilde{\omega}_p \sim \xi^{-1/2}$.
This dependence is shown in Fig.~\ref{Fig04}(b) by the broken red line, whereas it is beyond the limits of the frame in Fig.~\ref{Fig04}(a).

The estimate of the frequency downshift as per the approximate general relation (\ref{PeakFrequencyVariation_Effective}), which represents the linear trend valid for short distances, is plotted in Fig.~\ref{Fig04} with the black dashed-dotted curves. 
One may conclude that it gives accurate estimates for the peak frequency until the wave spectrum is not too broad. 
In the case shown in Fig.~\ref{Fig04}(a) the eventual frequency downshift is no more than just a few percent.

In the same Fig.~\ref{Fig04} we plot with the pink curve the function $\sqrt{S(\xi,\omega_0)/S_0(\omega_0)} = \exp{(-\xi)}$, which can be used for the rough estimation of the maximum spectral amplitude. 
Due to the frequency shift, the actual amplitude of the spectral peak is somewhat larger, see Fig.~\ref{Fig03}. 
Indeed, due to the frequency-dependent dissipation, amplitudes of lower Fourier harmonics decay slower than the higher ones. 
This is clear from Fig. \ref{Fig03} that the Fourier amplitude at $\omega_p$ for $t > 0$ is greater than at $\omega_0$. 
See also the difference between the pink and black lines in Fig. \ref{Fig04}(b).
Note that the dimensionless coordinate $\xi$ is normalized by the initial peak frequency $\omega_0$.
The relative spectral amplitude $\sqrt{S(\omega_p)/S_0(\omega_0)}$ 
according to Eq.~(\ref{SpeactralDecay_Effective}) and using the approximate solution (\ref{PeakFrequencyVariation_Effective}) is compared in Fig.~\ref{Fig04} with the exact solution (circles against the solid black line). The approximate estimate reproduces the exact solution with high accuracy in both cases shown in Fig.~\ref{Fig04}; it is noticeably different from  $\exp{(-\xi)}$ in the case of a short wavetrain with a broad spectrum (Fig.~\ref{Fig04}b).  

Owing to the exponentially fast decay of the spectral amplitudes, solutions shown in Fig.~\ref{Fig04} by pink lines become effectively zero when $\xi \gtrsim 6$.
Albeit according to Eqs.  (\ref{MeanFrequencyRate_NarrowBand}) and (\ref{PeakFrequencyVariation_Initial}) the dominant frequency formally decays to zero when $x \to \infty$, the practically distinguishable downshifting of the spectrum maximum occurs only within the characteristic width of the initial spectrum, because of the absence of the energy flux along the spectrum within the framework of the linear theory. 

Note that within the linear theory, the process of wavetrain fading process including the frequency shift does not depend on the particular form of the real part of the dispersion relation $k_r(\omega)$.

\section{THE DISSIPATIVE NONLINEAR SCHR\"ODINGER EQUATION FOR WEAKLY NONLINEAR MODULATED WAVES} 
%{PERTURBATIONS}
\label{Sect-3}

As the nonlinearity can result in the energy redistribution within the spectrum and, therefore, can influence the downshifting effect, we consider now a weakly nonlinear theory for narrow-band surface waves in a deep ocean covered by ice floes of the average thickness $h$; this will allow us to get some analytic results and insight into the complex physical phenomenon. 
The dispersion equation for such waves in  a perfect fluid is well-known (see, for example, Ref.~\onlinecite{Slunyaev-22} and references therein):
\begin{equation}
\label{E01}
\omega (k)= \sqrt{\frac{gk}{1 + Mk}},
\end{equation}
where the parameter $M = \rho_i h/\rho$ takes into account the floe inertia. 
This equation can be rewritten in the equivalent form
\begin{equation}
\label{E02}
k(\omega) = \frac{\omega^2}{g - M\omega^2},
\end{equation}
suitable for solving the boundary value problem when a wave motion is generated at a certain place of space $x_0=0$ as a function of time, and a perturbation further evolves along the $x$-axis. 

The weakly nonlinear evolution of slowly modulated waves can be described by the nonlinear Schr\"odinger equation [\onlinecite{Ablowitz-1981}], [\onlinecite{Osborne-2010}]. 
It is convenient to employ the generalized dissipative NLS equation in the form suitable for the description of the wave evolution along the $x$-axis, similar to that considered by Alberello \& P$\breve a$r$\breve a$u [\onlinecite{Alberello-22}]:
\begin{equation}
	\label{Eq01}
	i\left(\frac{\partial \psi}{\partial x} + \frac{1}{c_g}\frac{\partial \psi}{\partial t}\right) + \frac{\beta}{c_g^3}\frac{\partial^2 \psi}{\partial t^2} + \frac{\alpha}{c_g} |\psi|^2\psi = i \hat R(\omega)\psi.
\end{equation}
Here the surface displacement $\eta(x,t)$ is expressed through the complex wave envelope $\psi(x,t)$, $\eta = \text{Re}{\left[\psi e^{i(\omega_0 t - k_0x)} \right] }$.
The operator in the Fourier space $\hat{R}(\omega) = k_i(\omega)$ describes the dissipation effect through the effective imaginary part of the dispersion relation $k(\omega) = k_r(\omega) + i k_i(\omega)$. 
The NLS equation is valid for narrow-banded wave processes with the carrier frequency $\omega_0$ and the corresponding carrier wavenumber $k_0=k_r(\omega_0)$. 
While the real part of the complex dispersion relation $k(\omega)$ is given by Eq. (\ref{E02}), the NLS equation contains the approximate dispersion relation due to the narrowbandness assumption, $(k_r - k_0) \approx (\omega - \omega_0)/c_g + \beta (\omega-\omega_0)^2/c_g^3$. 
The group velocity $c_g$ as well as the nonlinear $\alpha$ and dispersive $\beta$ coefficients respectively are calculated for the carrier wave; their expressions can be found in Ref.~[\onlinecite{Slunyaev-22}]. 
In the deep-water limit, these parameters are:
\begin{equation}
\label{Eq-Nonlin}
c_g = \frac{\left(g - M\omega_0^2\right)^2}{2g\omega_0}, \quad \alpha = \frac{\omega_0^5}{4g^2}\frac{2g + M\omega_0^2}{g - M\omega_0^2}, \quad \beta = \frac{\left(g + 3M\omega_0^2\right)\left(g - M\omega_0^2\right)^3}{8g^2\omega_0^3}.
\end{equation}
Owing to the narrowband assumption, the operator $\hat{R}$ can be replaced by a few first terms of the Taylor series for the function $k_i(\omega)$ expanded in the vicinity of the carrier frequency  $\omega_0$.
The Taylor series expansion up to the therm $(\omega - \omega_0)$ inclusive gives:
\begin{equation}
\label{Eq01a}
\hat{R} = k_i \approx - \mu \left[\omega_0^n + n\omega_0^{n-1}(\omega - \omega_0)\right].
\end{equation}
Using the well-known rule for the correspondence between the dispersion relation and linear partial differential equation [\onlinecite{Korpel84}], $\omega - \omega_0 \to i\partial/\partial t$, we transform the approximate operator for dissipation (\ref{Eq01a}) to the corresponding terms of the equation, and write down the generalized NLS equation with the dissipative terms on the right-hand side:
\begin{widetext}
\begin{equation}
\label{Eq01b}
i\left(\frac{\partial \psi}{\partial x} + \frac{1}{c_g}\frac{\partial \psi}{\partial t}\right) + \frac{\beta}{c_g^3}\frac{\partial^2 \psi}{\partial t^2} + \frac{\alpha}{c_g} |\psi|^2\psi = -i\mu\omega_0^{n} \left(\psi + \frac{in}{\omega_0} \frac{\partial \psi}{\partial t}\right).
\end{equation}
\end{widetext}
Here, unlike in Ref. \onlinecite{Alberello-22}, the linear operator $\hat R(\omega)$ is presented by only two leading-order terms in the vicinity of the particular frequency $\omega_0$. 

The third-order asymptotic theory (\ref{Eq01b}) implies that the effect of a weak dispersion should be in balance with the small nonlinearity, $\left|\omega - \omega_0\right|/\omega_0 = O(\epsilon)$, that is characterized by the steepness $\epsilon = k_0 A \ll 1$, where $A$ is the typical amplitude of the water surface displacement.
Respectively, the parameter $\mu$ must be small to secure the weakness of dissipative terms in Eq. (\ref{Eq01b}).
As we will discuss further in Sec.~\ref{Sect-3}, the second term of the Taylor expansion (\ref{Eq01a}) (and, correspondingly, the second term in the right-hand side of Eq. (\ref{Eq01b})), supports an essentially nonlinear regime of the frequency downshifting, whereas this effect does not occur when the dissipation term is frequency independent.

\begin{figure}[h!]
%\vspace{-0.5cm}
\centerline{\includegraphics[width=1.1\textwidth]{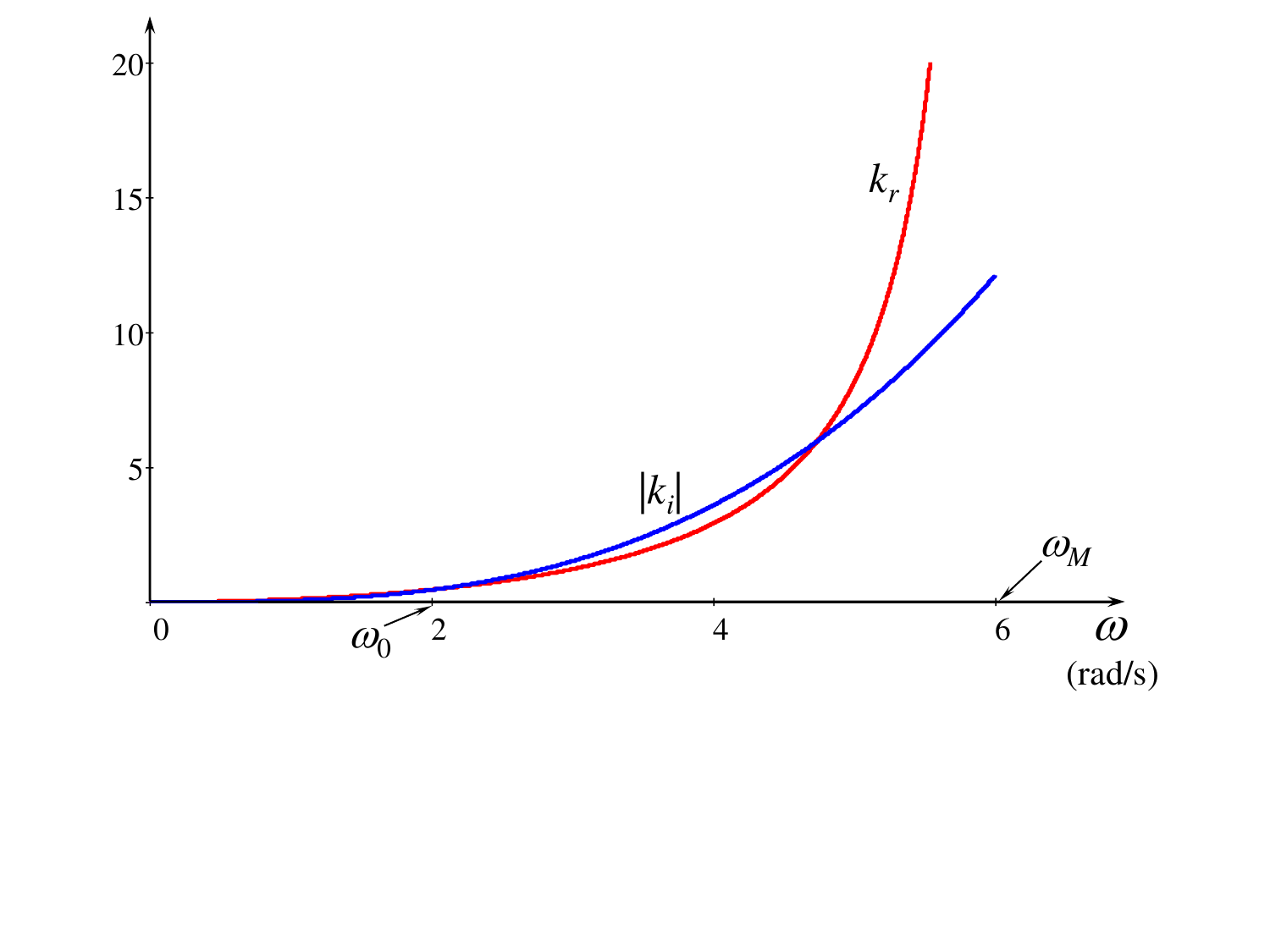}}%
\vspace{-3.5cm}
\caption{(Color online.) The real $k_r(\omega)$ and imaginary $k_i(\omega) = -\mu\omega^3$ parts of the dispersion relation of surface waves with floes for relatively weak dissipation with $\nu = 0.02$~s$^{-1}$. 
The real part is plotted according to Eq. (\ref{E02}); the imaginary part is multiplied by the factor $10^3$ to make it comparable with the real part. 
The dashed vertical line shows the limiting frequency $\omega_M \equiv \sqrt{g/M}$. 
The black solid line schematically shows the shape of the wavetrain with the carrier frequency $\omega_0 = 2$~rad/s ($\omega_0 T \approx 11.13$).}
\label{Fig05}
\end{figure}

The real and imaginary parts of the dispersion relation $k_r$ and $|k_i|$ are shown in Fig.~\ref{Fig05} for the particular case of the frequency-dependent dissipation with $n=3$ and the following set of parameters:
\begin{equation}
\label{param}
\omega_0 = 2\; \mbox{rad/sec}, \quad \rho = 1025 \; \mbox{kg/m}^3, \quad \rho_i = 922.5 \; \mbox{kg/m}^3, \quad h = 0.3 \; \mbox{m}, \quad \nu = 0.02\; \mbox{s}^{-1}.
\end{equation}
For the sake of inter-comparison, these parameters were chosen the same as in Ref.~[\onlinecite{Alberello-22}] for the ``weak dissipation case'' (in Ref.~[\onlinecite{Alberello-23}], another classification of weak and strong dissipation cases is suggested). 
This yields the dissipation coefficient $\mu = 5.6 \cdot 10^{-5}$~s$^3$/m. 
For the ``strong dissipation case'' with $\nu = 0.2 \; \mbox{s}^{-1}$ as per Ref.~[\onlinecite{Alberello-22}] and the same frequency $\omega_0$, we obtain $\mu = 5.6 \cdot 10^{-4}$~s$^3$/m. The range of applicability of the asymptotic theory will be discussed below.

%****************************************************************
%****************************************************************

In Ref.~\onlinecite{Alberello-22}, the influence of floe inertia was totally neglected which corresponds to the limit $M = 0$ in the formulae for coefficients (\ref{Eq-Nonlin}) and in the dispersion relations (\ref{E01}) and (\ref{E02}). 
It is clear from Eq. (\ref{E01}) that the effect of inertia becomes significant when $k_0 M = O(1)$, which happens when the dominant wavelength is comparable with the ice thickness.
The effect of ice inertia strengthens for shorter waves and dominates when $\omega_0$ approaches $\omega_M \equiv \sqrt{g/M}$, where $\omega_M \approx 6.2$~rad/s for the parameters listed in (\ref{param}). 
The proximity to this limit can be characterized by the small parameter $\gamma = (\omega_M - \omega_0)/\omega_0 \ll 1$. 
In this limit of short waves, the group velocity tends to zero, $c_g/(h \omega_M) = O(\gamma^2)$, and the coefficients of the nonlinear and dispersive terms in the NLS equation infinitely grow , $\alpha / c_g \propto O(\gamma^{-3})$, \; $\beta / c_g^3 \propto O(\gamma^{-1})$. 
%However, the coefficients of the TNLS equation remain finite after the following scaling of variables:
However, the NLS equation (\ref{Eq01b}) can be rescaled:
\begin{eqnarray}
\psi(x,t)=c_g \sqrt{\frac{2}{\alpha \beta} } \Psi(\chi,\tau), \quad x = \frac{\beta}{c_g}\chi, \quad t = \frac{\beta}{c_g^2} \tau
\end{eqnarray}
and presented in the dimensionless form:
\begin{widetext}
\begin{equation}
\label{TNLS-Dimensionless}
i\left(\frac{\partial \Psi}{\partial \chi} + \frac{\partial \Psi}{\partial \tau}\right) + \frac{\partial^2 \Psi}{\partial \tau^2} + 2 |\Psi|^2\Psi = -i \mu\omega_0^{n} \left( \frac{\beta}{c_g}  \Psi + \frac{in c_g}{\omega_0} \frac{\partial \Psi}{\partial \tau}\right).
\end{equation}
\end{widetext}
It now includes only two composite coefficients at the dissipation terms determined by the physical properties of the ice for the given carrier wave frequency. 
The right-hand side of Eq.~(\ref{TNLS-Dimensionless}) can be considered as the weak perturbation when the coefficient $\mu$ is sufficiently small. 
In the short-wave limit, the smallness of the dissipative terms is additionally secured by the small factors $\beta/c_g \propto \gamma$ and $c_g \propto \gamma^2$, which weakens the restriction on the smallness of the parameter $\mu$.

As one can see from Eq. (\ref{Eq-Nonlin}), the coefficients $\alpha$ and $\beta$ are both positive, therefore the NLS equation is of the focusing type that admits solutions in the form of envelope solitons [\onlinecite{Ablowitz-1981}], [\onlinecite{Osborne-2010}].
These nonlinear coherent structures are eternal within the NLS equation when the dissipation is absent ($\mu=0$), and can be long-lived in the real ocean [\onlinecite{Slunyaev2021}], [\onlinecite{Onoratoetal2021}] noticeably affecting the appearance of stochastic oceanic wave fields.
Following Fabrikant [\onlinecite{Fabrikant84}], [\onlinecite{Fabr&Step98}], below we consider the influence of dissipation on an envelope soliton in the course of its evolution.
	 
\section{ADIABATIC DECAY OF AN ENVELOPE SOLITON AND FREQUENCY DOWNSHIFTING}
\label{Sect-4}

The effect of the dissipative terms on the right-hand side of the NLS equation (\ref{Eq01b}) is presumed to be small compared with the dispersive and nonlinear terms on the left-hand side of this equation. This assumption allows us to consider a soliton solution to this equation with adiabatically slowly varying parameters (amplitude, width, speed, etc.).
When there is no dissipation, the NLS equation (\ref{Eq01b}) with $\mu = 0$ has an exact solution in the form of the envelope soliton with the carrier wavenumber which is connected with the frequency through the dispersion relation. The soliton solution can be presented as [\onlinecite{Grimshaw-16}], [\onlinecite{Stepanyants-19}]:
\begin{equation}%
\label{Eq02}%
\psi = A\sech{\left[\frac{1}{T}\left(t -\frac{x}{V}\right)\right]}e^{i(\sigma t - \kappa x)}.
\end{equation}
This solution contains two independent parameters which can be chosen arbitrarily. 
For our purposes, it is convenient to choose the soliton amplitude $A$ and the frequency shift $\sigma$ as the independent parameters. 
The perturbation of the water surface with the carrier wave of a frequency $\omega_0$ and envelope (\ref{Eq02}) is: 
\begin{equation}
\label{Surface}   
\eta = \text{Re}{\left[\psi e^{i(\omega_0 t - k_0x)} \right]} = A\sech{\left[\frac{1}{T}\left(t -\frac{x}{V}\right)\right]}\text{Re}{\left[e^{i(\omega_0 + \sigma)t - i(k_0 + \kappa)x}\right]}.
\end{equation}
So that the actual frequency is $\omega = \omega_0 + \sigma$ and the actial wavenumber is $k=k_0+\kappa$.
Then, the characteristic duration of soliton, $T$, its speed $V$, and the nonlinear correction to the wavenumber $\kappa$ can be presented in terms of $A$ and $\sigma$:
\begin{widetext}
\begin{equation}%
\label{Eq03}%
T = \frac{1}{Ac_g}\sqrt{\frac{2\beta}{\alpha}}, \quad V = \frac{c_g^3}{c_g^2 + 2\beta\sigma}, \quad \kappa = \frac{1}{c_g}\left(\sigma + \frac{\beta}{c_g^2}\sigma^2 - \frac{1}{2}\alpha A^2\right).
\end{equation}
\end{widetext}
For the soliton solution, the dispersive and nonlinear terms in the NLS equation (\ref{Eq01}) or (\ref{Eq01b}) with $\mu = 0$ are of the same order of magnitude and balance each other providing the existence of stationary moving wave group.

Envelope solitons were reproduced under laboratory conditions in Ref.~[\onlinecite{Slunyaevetal2013}] for the range of wave steepness $k_0A$ from moderate to the maximum, $k_0A \approx 0.3$ when local micro-breaking phenomena were observed near the wave crests. 
As the soliton duration $T$ is inversely proportional to its amplitude (see Eq.~(\ref{Eq03})), envelope solitons consisting of steeper waves are shorter. 
Considering for brevity the limit $M=0$, the following relation between the normalized envelope soliton amplitude and the soliton duration takes place, $\omega_0 T = \sqrt{2}/{(k_0A)}$. 
According to this formula, the choice $\omega_0 T = 4$ corresponds to the steepness $k_0 A\approx 0.35$ and thus can be considered as the narrowest realistic hydrodynamics envelope soliton. 
The other choice $\omega_0 T = 14$ yields $k_0 A\approx 0.10$ which is frequently considered as a typical steepness of nonlinear sea waves. 
Therefore, in this work we examine these examples as two representative cases; the corresponding surface displacements are plotted in Fig.~\ref{Fig02}.

When a small dissipation is taken into consideration, we assume the adiabatic regime of the wave evolution, so that the solution preserves the form of the envelope soliton described by Eq. (\ref{Eq02}) due to the balance between the effects of nonlinearity and dispersion, but with slowly varying in space parameters $A$ and $\sigma$ due to the dissipation. 
The weakness of the dissipation compared to the nonlinearity required for the validity of the nonlinear adiabatic approximation can be formally written as $\mu \omega_0^n \ll \alpha/c_g |A|^2$ (see Eq.~(\ref{Eq01b})), so that the soliton amplitude cannot be too small. 
This condition can be transformed to the form:
\begin{equation}%
\label{AdiabaticCondition}%
\mu \omega_0^n  \ll k_0 (k_0 A)^2 \left[ 1 + \frac{1}{2} \left( \frac{\omega_0}{\omega_M} \right)^2 \right],
\end{equation}
where the term in square brackets represents the contribution of the ice inertia; it varies from $1$ to $3/2$. 
The left-hand side of Eq.  (\ref{AdiabaticCondition}) represents the inverse characteristic scale of dissipation $\xi/x$, whereas the right-hand side is proportional to the inverse nonlinear spatial scale. 
This constraint can be re-written in terms of the wave amplitude:
\begin{equation}%
\label{AdiabaticCondition-Amplitude}%
A \gg A_l \equiv \sqrt{\frac{2\mu g^3 }{\omega_0^{6-n}}
\frac{\left( 1 - \omega_0/\omega_M\right)^3}{2 +  \left(\omega_0/\omega_M\right)^2}},
\end{equation}
which relaxes when the dissipation coefficient decreases.

There is also one more restriction on the wave amplitude that follows from the condition that the nonlinear term in Eq.~(\ref{Eq01b}) was assumed to be small in comparison with the advective term in the process of derivation of that equation.
This restriction leads to:
\begin{align} 
\label{SteepnessRestriction}
A \ll A_u \equiv \frac{2g}{\omega_0^2} \sqrt{\frac{1 -\left(\omega_0/\omega_M\right)^2}{2 + \left(\omega_0/\omega_M\right)^2}}.
\end{align}
In the limit when the ice cover is absent or insignificant (i.e., when $\omega_0 \ll \omega_M$), this gives $k_0 A \ll \sqrt{2}$.
In another limit of very strong ice inertia, the condition (\ref{SteepnessRestriction}) reduces to the restriction on the wave amplitude in the form, $A \ll M \sqrt{8\gamma/3}$. 
This implies that the wave amplitude ought to be much smaller than the ice thickness $h$, whereas the wave steepness $kA \ll \sqrt{2/3} \gamma^{-1/2}$ becomes unrestricted for small $\gamma$.

The ratio of the two limiting values for wave amplitudes is:
\begin{align} 
\label{LimAmpRatio}
\frac{A_u}{A_l} = \sqrt{\frac{2\omega_0^{2-n}}{\mu g}}\frac{\sqrt{1 + \omega_0/\omega_M}}{1 - \omega_0/\omega_M}.
\end{align}

\begin{figure}[h!]
\centerline{{\large (a)} \includegraphics[width=0.8\textwidth]{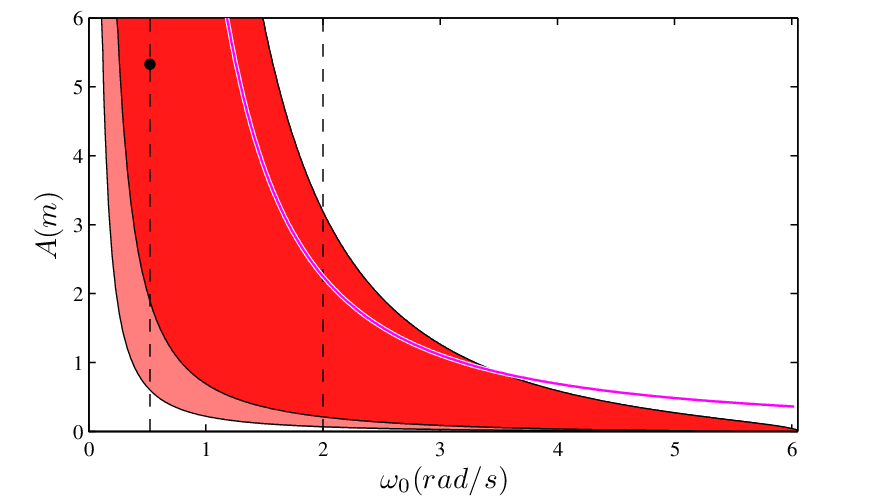}}
\centerline{{\large (b)} \includegraphics[width=0.8\textwidth]{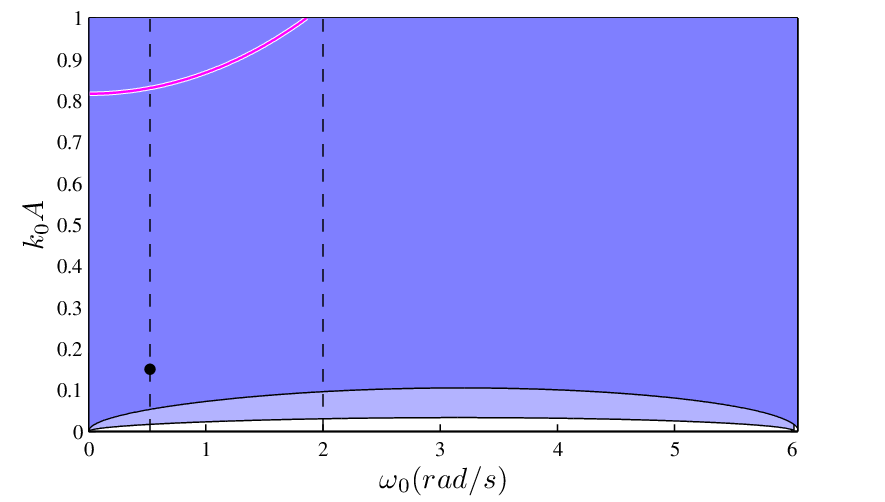}}
%\vspace{-0.5cm}
\caption{(Color online.) The limits of applicability for the wave amplitude (a) and steepness (b) as per Eqs.~(\ref{AdiabaticCondition-Amplitude}) and (\ref{SteepnessRestriction}) for the case $n=3$. 
The filling of any intensity corresponds to the weak dissipation case, while the deep-color filling -- to the strong dissipation case. 
The magenta curves correspond to the amplitude $A_c$ (\ref{Eq17}). 
The vertical dashed lines show frequencies $\omega_0 = 2\pi/12$~rad/s and $\omega_0 = 2$~rad/s.
The black circle corresponds to the parameters used in the numerical simulation of the NLS equation.}
\label{Fig06}
\end{figure}

The dependences (\ref{AdiabaticCondition-Amplitude}) and (\ref{SteepnessRestriction}) are displayed in Fig.~\ref{Fig06} for $n=3$ and other parameters presented in (\ref{param}). 
For these parameters Eq. (\ref{AdiabaticCondition-Amplitude}) gives $A \gg 0.6$~m  and $k_0A \gg 0.017$ in the weak dissipation case ($\nu = 0.02$~s$^{-1}$), and $A \gg 1.9$~m and $k_0A \gg 0.053$ in the strong dissipation case ($\nu = 0.2$~s$^{-1}$), see vertical dashed lines in Fig.~\ref{Fig06}. 
The quantity $A_u$ that specifies the upper boundaries in Fig.~\ref{Fig06} does not depend on the dissipation coefficient (the corresponding upper limit for the steepness is above the limit of the vertical axis in Fig.~\ref{Fig06}b). 
As discussed above, the range of applicability broadens to smaller wave amplitudes when the dissipation reduces.
In what follows, we consider two examples with the carrier frequency $\omega_0 = 2\pi/12$~rad/s and $\omega_0 = 2$~rad/s. 
The latter case admits waves of much smaller heights but it is more restrictive with respect to the wave steepness, $k_0A \gg 0.1$ when the dissipation is strong.
%****************************************************************
%****************************************************************

From the NLS equation (\ref{Eq01b}) one can derive several balance equations.
In particular, by multiplying Eq.~(\ref{Eq01b}) by the complex-conjugate function $\psi^*$, subtracting from the resultant equation its complex-conjugate counterpart and subsequent integration over $t$, we obtain: 
\begin{equation}
\label{Eq05}
\frac{dN}{dx} = -2\mu \omega_0^n \left(N - \frac{2n}{\omega_0} P\right).
\end{equation}
Another equation follows from Eq.~(\ref{Eq01b}) after multiplication by $\partial\psi^*/\partial t$, addition with the complex-conjugate equation, and subsequent integration over $t$:
\begin{equation}
\label{Eq06}
\frac{dP}{dx} = -\mu \omega_0^n \left(2 P - \frac{n}{\omega_0} D\right).
\end{equation}
Here we denote:
%\begin{widetext}
\begin{equation}
\label{Eq07}
N = \int\limits_{-\infty}^{+\infty}|\psi|^2dt, \quad
P = \frac{1}{2i}\int\limits_{-\infty}^{+\infty}\psi^*\frac{\partial \psi}{\partial t}\,dt, \quad
D = \int\limits_{-\infty}^{+\infty}\left|\frac{\partial \psi}{\partial t}\right|^2dt,
\end{equation}
and presume that the solution decays to zero when $t \to \pm \infty$.
If $\mu = 0$, then $N$ and $P$ are conserved quantities that do not depend on $x$.

To describe the variation of soliton parameters $A$ and $\sigma$ with $x$ one can apply the asymptotic theory (see, for example, [\onlinecite{Gorshkov81}], [\onlinecite{Ostrovsky-2022}]) provided that the dissipative terms on the right-hand side of the NLS equation (\ref{Eq01b}) are small compared to the nonlinear and dispersive terms on the left-hand side \cite{Fabrikant84,Fabr&Step98}.
The consecutive application of the asymptotic theory reduces to the exploitation of the balance equations (\ref{Eq05}) and (\ref{Eq06}); this approach will be elaborated below.

\subsection{The standard adiabatic approach}
\label{Subsect-3.1}

%To be specific, hereafter we consider the wave dynamics for the particular choice of the dissipation power law with $n=3$.
Assuming that in all expressions (\ref{Eq07}) the function $\psi$ describes an envelope soliton as per Eq. (\ref{Eq02}) with $x$-dependent parameters $A(x)$ and $\sigma(x)$ that slowly vary in space, and the remaining soliton parameters are specified by Eq.~(\ref{Eq03}), we obtain after simple manipulations:
%\begin{widetext}
\begin{equation}
\label{Eq08}
N(x) = 2\sqrt{\frac{2\beta}{\alpha}}\frac{A(x)}{c_g}, \quad 
P(x) = -N(x)\frac{\sigma(x)}{2}, \quad D = N(x)\left[\frac{\alpha}{6\beta}A^2(x)c_g^2 + \sigma^2(x)\right].%    
\end{equation}
%\end{widetext}
Substituting these quantities in Eqs. (\ref{Eq05}), we arrive at the set of equations for $A(\xi)$ and $\sigma(\xi)$:
%\begin{widetext}
%\begin{eqnarray}
%\frac{dA}{dx} &=& - 2\mu \omega_0^3 A \left(1 + 3\frac{\sigma}{\omega_0}\right), \label{Eq11}\\
%\frac{d\sigma}{dx} &=& - \mu \omega_0^2c_g^2\frac{\alpha}{\beta} A^2. \label{Eq12}
%\end{eqnarray}
%
\begin{eqnarray}
\frac{dA}{d\xi} &=& - 2 A \left(1 + n\frac{\sigma}{\omega_0}\right),
\label{Eq11}\\
\frac{d\sigma}{d \xi} &=& - \frac{n}{3}  \frac{c_g^2 \alpha}{\omega_0 \beta} A^2, \label{Eq12}
\end{eqnarray}
%\end{widetext}
where $\xi = \mu \omega_0^n x$ is the normalized spatial coordinate as before. 
Note that the right-hand side of Eq.~(\ref{Eq12}) is negative which means that the frequency shift $\sigma$ always decreases with the distance, i.e., a permanent frequency downshifting occurs.
Applicability of Eq. (\ref{Eq11}) presumes that $-1/n < \sigma/\omega_0 < 0$; otherwise, it suggests a solution that is growing with $x$.
This condition implies that the second term on the right-hand side of Eq.~(\ref{Eq01b}) should remain smaller than the first one.

For the comparison, let us consider the frequency shift within the linear theory for a wavetrain of the same initial shape as the envelope soliton. 
For the early stage of the wavetrain evolution, the general formula (\ref{PeakFrequencyVariation_Initial}) can be reduced to the form:
\begin{equation} 
\label{PeakFrequencyVariation_Sech}
\frac{d \sigma}{d\xi} \Big|_{lin} \approx - 2n \omega_0 \delta_p^2.
\end{equation}%
The spectral width parameter $\delta_p$ for the pulse of the envelope soliton shape was calculated in Sec.~\ref{sec:ExampleLinearSech} where it was found that $\delta_p^2 =2 \left( \pi \omega_0 T \right)^{-2}$. 
Using Eq.~(\ref{Eq03}), we can relate the pulse duration $T$ with its amplitude $A$. 
Then, the equation (\ref{Eq12}) that is derived within the nonlinear asymptotic theory can be presented in the form:
\begin{equation} 
\label{Eq12_AlternativeForm}
\frac{d \sigma}{d\xi} = - \frac{\pi^2}{3}n \omega_0 \delta_p^2.
\end{equation}%
This expression looks similar to Eq.~(\ref{PeakFrequencyVariation_Sech}) up to the coefficient $\pi^2/3 \approx 3.3$ instead of $2$ on the right-hand side.
Hence, according to these estimates, a linear wavetrain of a soliton shape has a slower rate of frequency downshifting with the distance in comparison with the nonlinear envelope soliton.

\begin{figure}[h!]
%\vspace{-0.5cm}
\centerline{\includegraphics[width=0.85\textwidth]{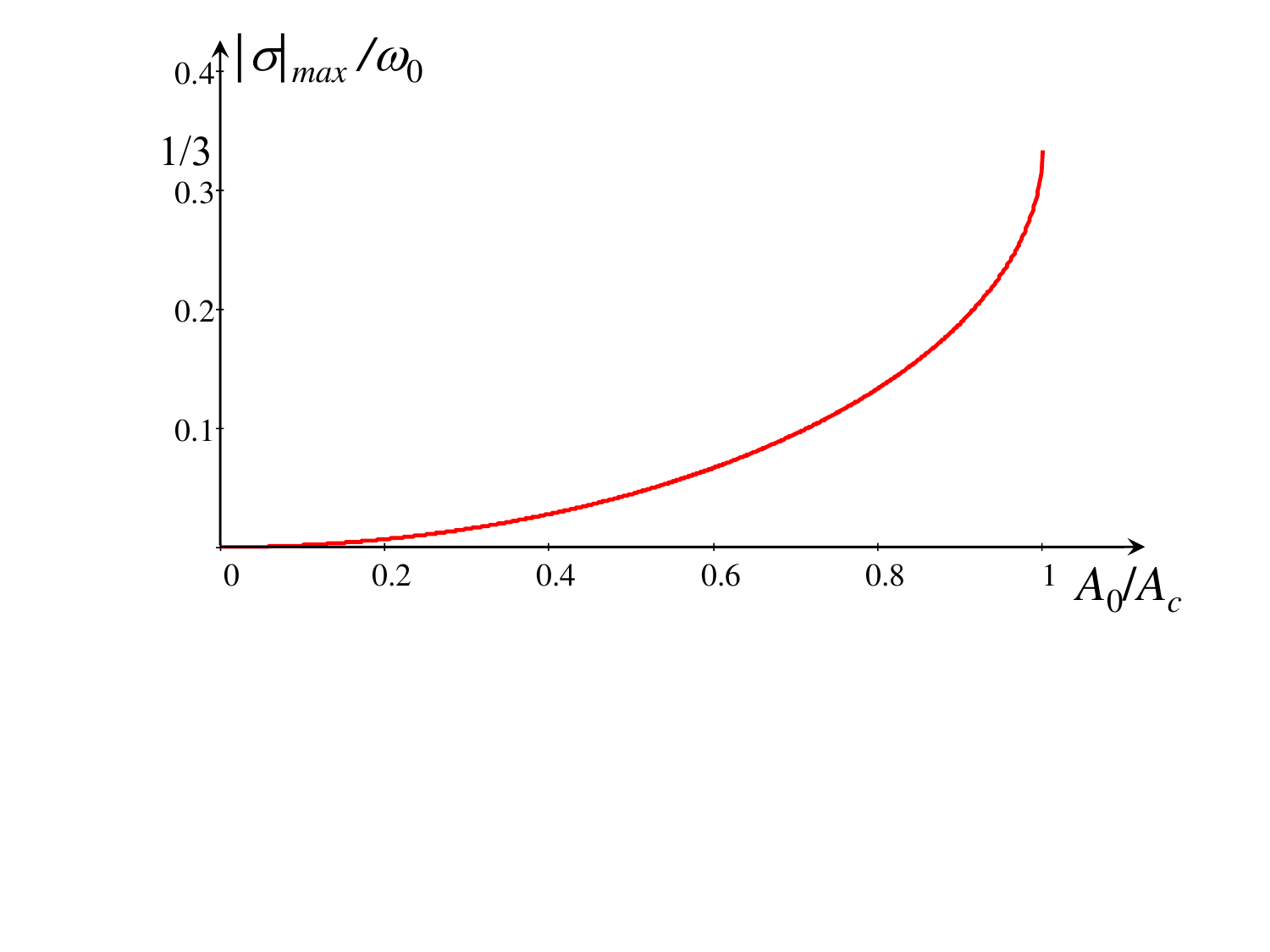}}%
\vspace{-4.0cm}
\caption{(Color online.) The normalised frequency shift $|\sigma|_{max}/\omega_0$ as the function of normalized amplitude $A_0/A_c$ as per Eq.~(\ref{sigma_max}) at $n = 3$. }
\label{Fig08}
\end{figure}

Dividing Eq.~(\ref{Eq11}) by Eq.~(\ref{Eq12}), we get:
\begin{equation}%
\label{Eq16}%
\frac{dA}{d\sigma} = \frac{6}{n} \frac{\beta\omega_0}{c_g^2\alpha A} \left(1 + n\frac{\sigma}{\omega_0}\right).
\end{equation}
By integration of this equation, we derive the relationship between $A$ and $\sigma$ which is the first integral of the system (\ref{Eq11}) \& (\ref{Eq12}):
\begin{align}%
\label{Eq17}%
\frac{A^2}{A_c^2} = \frac{A_0^2}{A_c^2} \left(1 + n \frac{\sigma}{\omega_0} \right)^2 -1, \quad A_c \equiv\sqrt{\frac{6}{n^2} \frac{\beta\omega_0^2}{c_g^2\alpha}} = \frac{2\sqrt{3}g}{n \omega_0^2} \sqrt{\frac{1 +3\left(\omega_0/\omega_M\right)^2}{2 + \left(\omega_0/\omega_M\right)^2}}, 
\end{align}
where $A_0$ is the initial amplitude of the envelope soliton amplitude $A(\xi)$ at $\xi = 0$ when $\sigma = 0$. 
Note that this solution does not depend on the dissipation parameter $\mu$; it is universal as long as the wavetrain evolution is adiabatic. 
The right-hand side of Eq.~(\ref{Eq17}) is a quadratic function of the parameter $\sigma$ that has a formal minimum at $\sigma/\omega_0 = -1/n$. 
Recall that such a frequency shift is beyond the range of applicability of the theory. 
When the initial soliton amplitude $A_0$ is such that $A_0 \le A_c(\omega_0)$
%
%\begin{align} 
%A_0 \le A_c(\omega_0) \equiv\sqrt{\frac{6}{n^2} \frac{\beta\omega_0^2}{c_g^2\alpha}} = \frac{2\sqrt{3}g}{n \omega_0^2} \sqrt{\frac{1 +3\left(\omega_0/\omega_M\right)^2}{2 + \left(\omega_0/\omega_M\right)^2}}, \\
%\end{align}
%
and the frequency shift decreases below the zero, then the amplitude $A$ monotonically decreases below $A_0$ and turns to zero at some point $-1/n \le \sigma_{min}/\omega_0 < 0$, where
\begin{align} 
\label{sigma_max}
\frac{\sigma_{min}}{\omega_0} = \frac{1}{n} \left[\sqrt{1- \left( \frac{A_0}{A_c} \right)^2} - 1 \right].
\end{align}
This formula shows that for a soliton of infinitesimal amplitude ($A_0 \to 0$), the maximal frequency shift is $|\sigma|_{max} \approx \left(\omega_0/2n\right)A_0^2/A_c^2$.
In another limit, $A_0 \to A_{cr}$, the maximal frequency shift is quite notable, $|\sigma|_{max} = \omega_0/n$.
Figure \ref{Fig08} shows the normalised frequency shift $|\sigma|_{max}/\omega_0$ as the function of normalized amplitude $A_0/A_c$ as per Eq.~(\ref{sigma_max}).
%****************************************************************
%**************************************************************

\begin{figure}[h!]
%\vspace{-0.5cm}
\centerline{\includegraphics[width=1.0\textwidth]{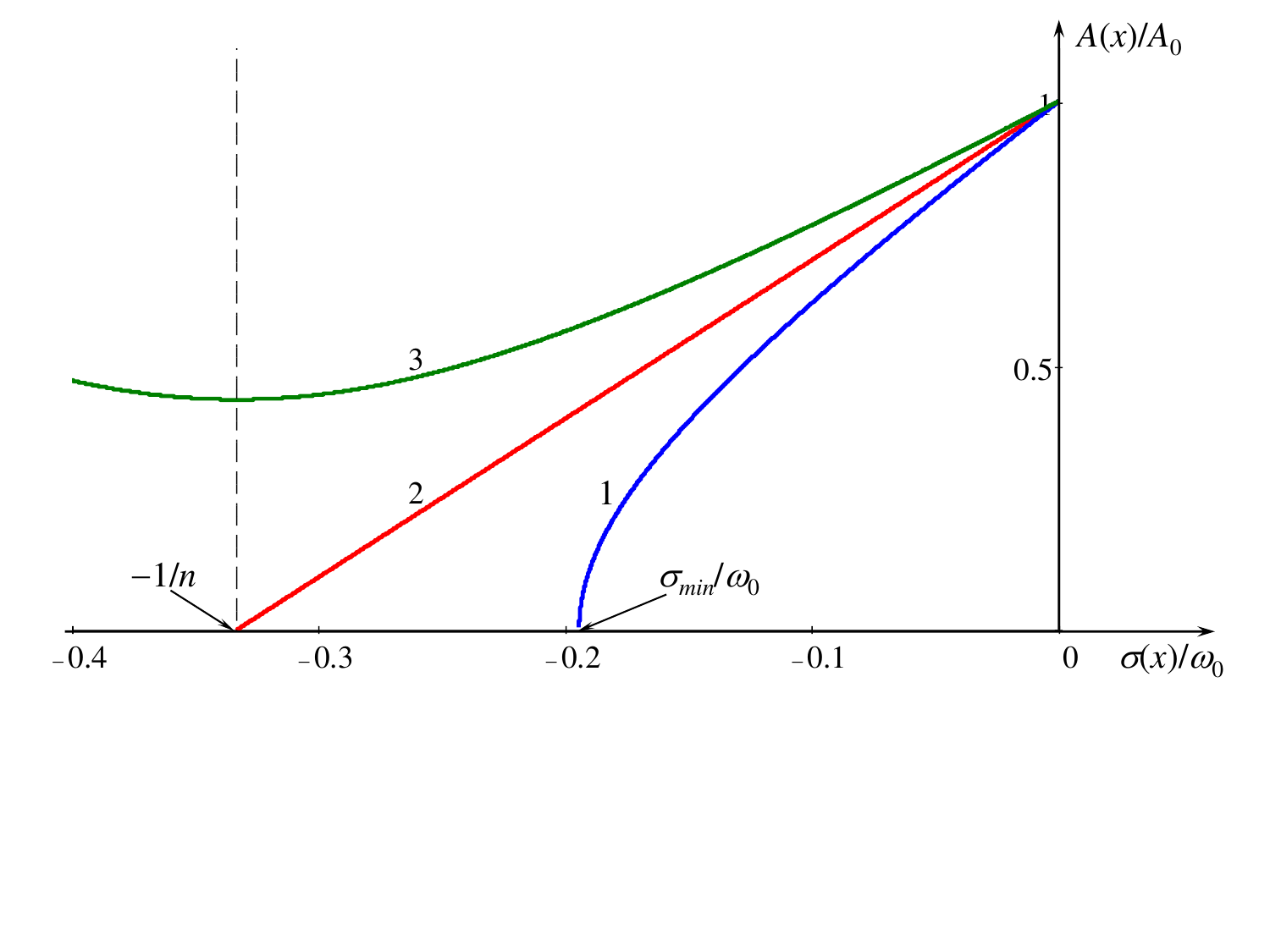}}%
\vspace{-3.5cm}
\caption{(Color online.) The dependence of $A/A_0$ on $\sigma/\omega_0$ as per Eq.~(\ref{Eq17}) for $n = 3$ and three different values of $A_c/A_0$. Line 1 pertains to $A_0 < A_c$, line 1 pertains to $A_0 = A_c$, and line 3 pertains to $A_0 > A_c$.}
\label{Fig15}
\end{figure}

Figure \ref{Fig15} illustrates the dependences $A(\sigma)$ for three typical cases of initial soliton amplitude for $n = 3$.
In particular, when $A \le A_c$, the final downshifting monotonically increases in magnitude when $A_0$ grows from zero to the critical amplitude $A_c(\omega_0)$. 

%****************************************************************
%****************************************************************
In the case of $A_0 = A_{c}$ the dependence of $A/A_0$ on $\sigma/\omega_0$ becomes very simple, $A/A_0 = 1 + n\sigma/\omega_0$ -- see line 2 in Fig.~\ref{Fig15}. 
In such a case, the frequency shift formally attains a quite notable value $\sigma = -\omega_0/n$.

In the third case when $A_0 > A_c$, the dependence of $A/A_0$ on $\sigma/\omega_0$ is shown in Fig.~\ref{Fig15} by line 3. 
In this case, the soliton amplitude achieves minimum value at $\sigma/\omega_0 = -1/n$, and then it formally increases when $\sigma/\omega_0$ continues decreasing. 
However, solutions within the framework of asymptotic theory become invalid when $\sigma/\omega_0$ approaches the value of $-1/n$ from above.

The relationship between the upper allowed amplitude $A_u$ as given in Eq.~(\ref{SteepnessRestriction}) and the critical amplitude $A_c$ is as follows:
\begin{align} 
\label{AuVsAc}
\frac{A_u}{A_c} = \frac{n}{\sqrt{3}} \sqrt{\frac{1 -\left(\omega_0/\omega_M\right)^2}{1 + 3 \left(\omega_0/\omega_M\right)^2}}.
\end{align}
The condition $A=A_c$ is shown in Fig.~\ref{Fig06} by magenta lines. 
The expression (\ref{AuVsAc}) implies that $A_c \ge A_u\sqrt{3}/n$; therefore, the amplitude $A_0 > A_c$ may fall within the formal limit of applicability of the weakly nonlinear NLS equation only if the power $n$ is not small. 
At the same time, the corresponding steepness (the area above the magenta curve in Fig.~\ref{Fig06} for the case $n=3$) is physically unrealistic. 
It is constructive to reformulate the condition for $A_c$ in terms of the soliton width using Eq.~(\ref{Eq03}); then it reads $T_0 \ge T_c$, where $\omega_0 T_c = n/\sqrt{3}$, and $T_0$ is the initial envelope-soliton duration. 
Therefore, within the adiabatic theory, only unrealistically short envelope solitons can experience downshifting below the critical frequency $\omega_0/n$ where soliton amplitude formally begins to grow with the distance.

From this analysis, we can conclude that under reasonable conditions the frequency downshift accrued as a result of the adiabatic nonlinear wave transformation cannot be greater than $\omega_0/n$. This means that line 1 in Fig.~\ref{Fig15} represents the general case feasible in application to oceanic water waves.
As was mentioned above, for small-amplitude solitons the dissipation cannot be considered weak (see Eq.~(\ref{AdiabaticCondition-Amplitude})); therefore, a deviation from the adiabatic theory can be expected at the late stage of soliton evolution.

If the frequency shift $|\sigma|/\omega_0$ is small (for example, at the initial stage of soliton evolution), then the solution (\ref{Eq17}) can be approximately presented as:
\begin{equation}%
\label{Eq17-Approximate}%
\left(\frac{A}{A_0}\right)^2 \approx 1 + \frac{12}{n} \frac{\beta\omega_0^2}{c_g^2\alpha A_0^2} \frac{\sigma}{\omega_0} = 1 + 6 \left( \omega_0 T_0 \right)^2 \frac{\sigma}{\omega_0}. 
\end{equation}

In the meantime, since the dissipation coefficient $\mu$ enters the expression for the Fourier transform evolution (\ref{Eq26}) solely through the scaled distance $\xi$, then the relationship between $A$ and $\sigma$ is independent on the parameter $\mu$ within the linear theory too:
\begin{equation}%
\label{AmplitudeVsSigma-Linear}%
\frac{S(\omega_p)}{S_0(\omega_0)}\Big|_{lin} = \sech^2{\left[\frac{\pi}{2} \omega_0 T_0\frac{\sigma}{\omega_0}\right]}
\exp{\left[ \frac{\pi}{n}\omega_0 T_0 \left( 1 + \frac{\sigma}{\omega_0}\right) \tanh{\left(\frac{\pi}{2} \omega_0 T_0\frac{\sigma}{\omega_0}\right)} \right]}.
\end{equation}
Expanding this solution for small $\sigma/\omega_0$ and relating the spectral width parameter $\omega_0 T_0$ to the initial soliton amplitude according to Eq. (\ref{Eq03}), we obtain the expression of the same form as Eq.~(\ref{Eq17-Approximate}), but within the linear theory:
\begin{equation}%
\label{Eq17-LinearAnalogue}%
\frac{S(\omega_p)}{S_0(\omega_0)} \Big|_{lin} \approx 1 + \frac{\pi^2}{2n} \left( \omega_0 T_0 \right)^2 \frac{\sigma}{\omega_0}
= 1 + \frac{\pi^2}{n} \frac{\beta\omega_0^2}{c_g^2\alpha A_0^2} \frac{\sigma}{\omega_0}. 
\end{equation}
Comparing the numerical coefficients 12 and $\pi^2 \approx 9.87$ in Eqs. (\ref{Eq17-Approximate}) and (\ref{Eq17-LinearAnalogue}), respectively, we conclude that the amplitude of the wavetrain always decreases with $|\sigma|$ a little bit slower within the linear theory. 
Therefore, the inverse function characterizing the frequency shift $|\sigma|$ as a function of the wave amplitude increases always faster (being negative) within the linear theory in comparison with the nonlinear one. 
Here we associate the decay of the spectral amplitude $\left[ S(x,\omega_p)/S_0(\omega_0) \right]^{1/2}$ with the decay rate of the wavetrain amplitude $A(x)/A(0)$, which is true only within the adiabatic stage of soliton evolution, when the wavetrain is self-similar. 

%****************************************************************
%\begin{figure}[b!]
%%\vspace{-0.5cm}
%\centerline{\includegraphics[width=1.0\textwidth]{Fig07.pdf}}%
%\vspace{-3.5cm}
%\caption{(Color online.) The dependence of $A/A_0$ on $\sigma/\omega_0$ within the linear (line 1), asymptotic (line 2), and advanced asymptotic (line 3) theories.}
%\label{Fig07}
%\end{figure}
%****************************************************************

As follows from Eq. (\ref{Eq11}), at the early stage of evolution when $|\sigma/\omega_0| \ll 1$ and the second term in the relation can be neglected, $dA/d\xi \approx -2 A$, and therefore, the spatial decay rate of soliton amplitude is two times greater than the linear theory predicts. 
This is a consequence of the relationship between the soliton duration and amplitude, $T(\xi) \sim A(\xi)^{-1}$, whereas in the linear theory, such a dependence is absent.

The dependences $A(\xi)$ and $\sigma(\xi)$ can be obtained from Eqs.~(\ref{Eq11}) and (\ref{Eq12}) in the analytic form. 
Eliminating $A$ from Eq. (\ref{Eq12}) with the help of Eq. (\ref{Eq16}), we obtain: 
\begin{equation}
\label{Eq160}%
\frac{1}{2}\frac{d}{d\xi} \left(1 + \frac{n \sigma}{\omega_0}\right) = 1 - \left(\frac{A_0}{A_c} \right)^2 - \left(1 + \frac{n \sigma}{\omega_0}\right)^2.
\end{equation}
This equation can be readily solved by the separation of variables. 
The solution depends on the relationship between $A_0$ and $A_c$ which determines the sign of the term in the square brackets. 
In the most realistic case $A_0 < A_c$ (see line line 1 in Fig.~\ref{Fig15}), the solution of Eq.~(\ref{Eq160}) reads: 
\begin{equation}%
\label{FrequencyShiftVsDistance_Small_A0}%
\frac{n \sigma}{\omega_0} = a\frac{\left(a + 1\right) + \left(a - 1\right)e^{-4a\xi}}{\left(a + 1\right) - \left(a - 1\right)e^{-4a\xi}} - 1, \quad \mbox{where} \quad a^2 = 1 - \left( \frac{A_0}{A_c} \right)^2 > 0.
\end{equation}

If $A_0 > A_c$, then the solution to Eq.~(\ref{Eq160}) is:
\begin{align}%
%\label{Eq160}%
\frac{n \sigma}{\omega_0} = 
-b\tan{\left(2b\xi - \arccot{b}\right)} - 1, \quad \mbox{where} \quad b^2 = \left( \frac{A_0}{A_c} \right)^2 - 1 > 0.
\end{align}
This solution breaks when $\xi \to \xi_d$ where
\begin{align}%
\label{xi_TheoryBreakdown}
\xi_d = \frac{\arccot{b}}{2b} =
\frac{1}{2} \left( A_0^2/A_c^2-1 \right)^{-1/2} \arcsin{\frac{A_c}{A_0}} 
\end{align}
when the asymptotic theory becomes invalid (see line 3 in Fig.~\ref{Fig15}). 
After this distance, the solution formally describes decreasing frequency shift up to minus infinity when $\xi \to \xi_{max}$, where $\xi_{max} = \xi_d + (\pi/4) \left( A_0^2/A_c^2-1 \right)^{-1/2}$.
%We reiterate that the NLS theory is valid only for the distances well before the one given in (\ref{xi_TheoryBreakdown}), $\xi \ll \xi_d$.

In the limiting case $A_0=A_c$ (see line 2 in Fig.~\ref{Fig15}), the solution is represented by the simple algebraic dependence:
\begin{align}%
%\label{Eq160}%
\frac{n \sigma}{\omega_0} = \frac{1}{1+ 2\xi} - 1,
\end{align}
and yields $\sigma \to - \omega_0/n$ when $\xi \to \infty$.

Using now Eq.~(\ref{Eq17}), we can determine the dependence of soliton amplitude on the distance, $A(\xi)$. 
For small-amplitude solitons $A_0<A_c$, the solution reads:
\begin{align}%
%\label{Eq160}%
A(\xi) = A_0 \frac{\sqrt{A_c^2/A_0^2-1}}{\sinh{\left[ \arccosh{\left(A_c/A_0\right)} + 2\xi \sqrt{1-A_0^2/A_c^2}  \right]}}.
\end{align}
In the opposite case $A_0>A_c$ the solution is:
\begin{align}%
%\label{Eq160}%
A(\xi) = A_0 \frac{\sqrt{1-A_c^2/A_0^2}}{\cos{\left[ \arcsin{\left(A_c/A_0\right)} - 2\xi \sqrt{A_0^2/A_c^2-1} \right]}},
\end{align}
and in the limiting case $A_0=A_c$ the dependence $A(\xi)$ is algebraic:
\begin{align}%
%\label{Eq160}%
A(\xi) = \frac{A_0}{1+2\xi}.
\end{align}

To conclude this subsection, we remind that the NLS equation is applicable to wavetrains of narrow-band spectra in the vicinity of a certain frequency $\omega_0$.
Therefore, within the framework of the developed model, the frequency downshifting should not be too big.
This implies that must be $|\sigma| \ll \omega_0$; however, there are many examples when approximate formulae provide reasonable dependences even beyond the formal range of their applicability.
When the frequency downshifting becomes relatively big, all coefficients of the NLS equation, including the group speed, change, and solutions should be updated.
In the next subsection \ref{Subsect-3.2}, we present the advanced adiabatic approach to soliton solutions which allows us to construct an approximate solution with a relatively big frequency variation. 

\subsection{Advanced adiabatic approach}
\label{Subsect-3.2}

The approximate NLS equation (\ref{Eq01b}) is most accurate when its coefficients are calculated for the actual dominant wave frequency. 
Therefore, the theory can be improved if the coefficients of the NLS equation (\ref{Eq01b}) are taken frequency-dependent to account for the frequency downshifting effect.
The set of equations (\ref{Eq11}), (\ref{Eq12}) with varying $\omega_0$ and frequency dependent coefficients $\alpha$, $\beta$, and $c_g$ becomes more complex and not solvable analytically; however, it can be easily solved numerically.
The results of calculations show that the dependences of the amplitude and frequency on the distance remain qualitatively the same as in the case with the constant-coefficient NLS equation but the amplitude decreases a bit slower, and the total frequency downshifting becomes lesser.
This can be explained by the strong dependence of the decay rate on the frequency $k_i \sim \omega^3$.
When the frequency shifts down, the amplitude decay rate quickly diminishes.
At a small distance, the result of the simplified theory and numerical data coincide.

\section{Results of the direct numerical simulation and comparison with the theory}
\label{Sect-5}

The direct numerical simulation of the NLS equation (\ref{Eq01b}) was performed using a standard pseudo-spectral technique. 
The evolution of the nonlinear part was obtained by integration using the 4-order Rounge--Kutta method with the adaptive spatial step size, while the linear part was evaluated at each step using the exact solution in the Fourier domain.

\begin{figure}[h!]
\centerline{\includegraphics[width=0.9\textwidth]{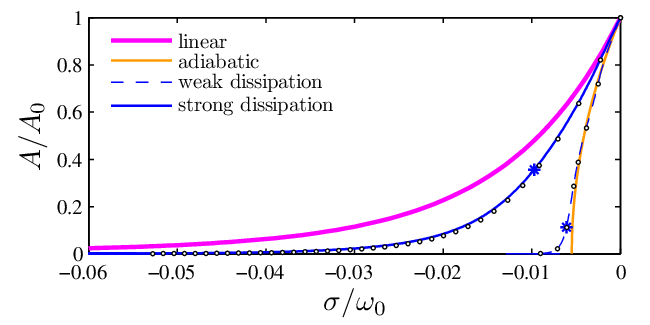}}%
\vspace{-0.5cm}
\caption{(Color online.) Relationships between the wavetrain amplitude $A/A_0$ and frequency downshift $\sigma/\omega_0$ that follow from different theories. 
The thick pink line pertains to the linear theory (\ref{AmplitudeVsSigma-Linear}), the solid orange line pertains to the analytic adiabatic theory (\ref{Eq17}), results of direct numerical simulation of the NLS equation with the weak dissipation are shown by the blue dashed line, and with the strong dissipation -- by the blue solid line. 
The initial conditions in all cases were $\omega_0 = \pi/6$~rad/s and $k_0A_0=0.15$. The blue stars correspond to the amplitude limits $A=A_l$ when the adiabatic theory ceases to work according to Eq. (\ref{AdiabaticCondition-Amplitude}). 
Small empty circles correspond to the numerical data when the non-expanded dissipative term $\hat{R} = -\mu \omega^3$ in the NLS equation was used as in Ref.~\onlinecite{Alberello-22}.}
\label{Fig12}
\end{figure}

\begin{figure}[t!]
%\centerline{{\large(a)} 
%\includegraphics[width=0.7\textwidth]{Fig13a.eps}} 
%\centerline{{\large(b)} \includegraphics[width=0.7\textwidth]{Fig13b.eps}}%
%subfigure[]{
\includegraphics[width=11.8cm]{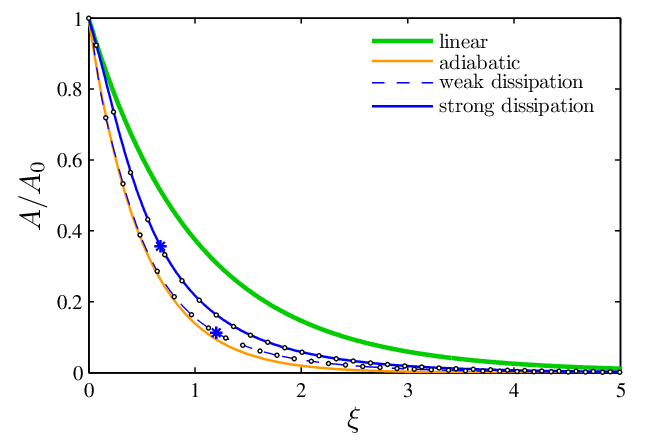} 
%}
%\subfigure[]{
\includegraphics[width=11.8cm]{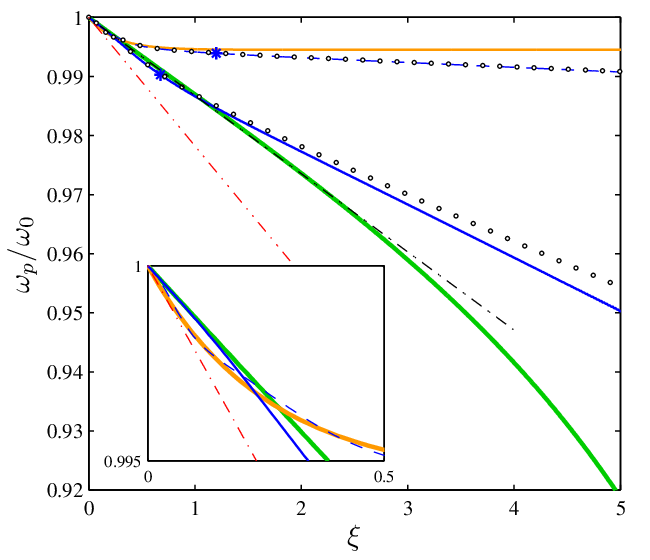}
%}
\vspace{-0.7cm}
\begin{picture}(300,0)%
\put(-40,430){{\Large a)}}%
\put(-40,175){{\Large b)}}%
\end{picture}
\caption{(Color online.) Dependences of the normalized soliton amplitude $A(x)/A_0$ (a) and peak frequency $\omega_p/\omega_0 = 1 + \sigma(x)/\omega_0$ (b) on the normalized distance $\xi = \mu\omega_0^3 x$ for the same cases as in Fig.~\ref{Fig12}. 
Stars correspond to the limiting amplitudes $A=A_l$. 
Trends of frequencies for small $|\xi|$ according to Eqs. (\ref{Eq12}) and (\ref{PeakFrequencyVariation_Sech}) are given in panel (b) by the red and black dashed-dotted lines, correspondingly.
The early stage of the frequency evolution is given in the insertion of panel (b).}
\label{Fig13}
\end{figure}
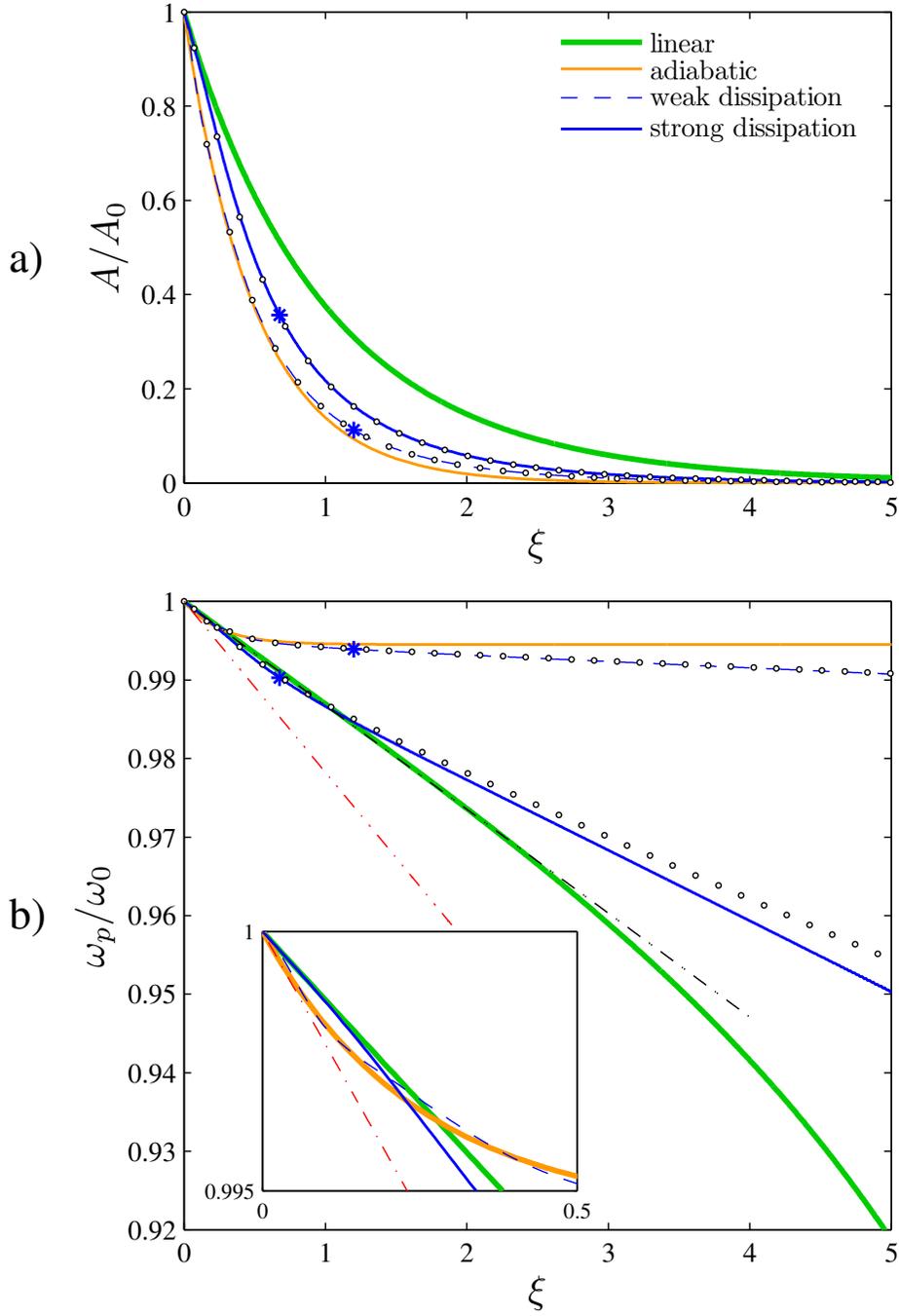

The simulations of decaying envelope solitons were performed for the particular power-type dissipation with $n=3$ and other parameters given in (\ref{param}) for the initial steepness of the soliton $k_0A_0 = 0.15$, $A_0\approx 5.3$~m (these conditions are marked in Fig.~\ref{Fig06} by filled circles) for two coefficients of dissipation corresponding to the ``weak'' ($\nu=0.02$ s$^{-1}$) and ``strong'' ($\nu=0.2$ s$^{-1}$) dissipation cases. 
The chosen wavetrain parameters correspond to the dimensionless wavetrain duration $\omega_0T_0\approx 9.6$ which represents an intermediate case between the two examples shown in Fig.~\ref{Fig02}.
Results of simulations are compared in Figs.~\ref{Fig12} and \ref{Fig13} with the linear theory for the wavetrain of the same shape (assuming that $A(\xi)/A_0 = \left[S(\omega_p(\xi))/S_0(\omega_0) \right]^{1/2}  $), and with the adiabatic theory for the NLS equation. 
To compare our results with the outcomes of  Ref.~\onlinecite{Alberello-22}, we undertook also a numerical simulation of the NLS equation taking in Eq. (\ref{Eq01}) the dissipative term in the full form $\hat{R}(\omega)=-\mu \omega^3$ without its expansion in the vicinity of $\omega = \omega_0$.
%***********************************

The relationship between the wavetrain amplitude and frequency downshift is shown in  Fig.~\ref{Fig12}. 
As was discussed above, the curves of the linear solution (\ref{AmplitudeVsSigma-Linear}) and of the adiabatic solution (\ref{Eq17}) do not depend on the strength of the dissipation.
It is clearly seen that in the case of a weak dissipation, the adiabatic theory (orange curve) agrees well with the direct numerical modelling of the NLS equation (blue dashed curve) until the wavetrain amplitude becomes of one order of magnitude smaller than initially. 
For smaller wave amplitudes the adiabatic solution departs from the numerical data and predicts a smaller total downshifting. 
In the case of a strong dissipation, the numerical data (see blue solid line in Fig.~\ref{Fig12}) are located between the linear and the adiabatic theories. 
Stars on the lines that correspond to the direct numerical simulation denote the critical amplitudes $A_l$ (\ref{AdiabaticCondition-Amplitude}) below which the dissipation term in the evolution equation prevails over the nonlinear term. One may conclude that the threshold criterion $A=A_l$ indeed gives a reasonable estimation when the adiabatic theory ceases to work in the case of a weak dissipation but has little sense in the case of relatively strong dissipation. 
The results of simulations with the dissipation term in the form of Ref.~\onlinecite{Alberello-22} are shown with empty circles which practically coincide with the simulations of the NLS equation in the form of Eq.~(\ref{Eq01b}). 
This can be explained by the narrow-band spectrum of the chosen initial wavetrain and a small downshift of the frequency in the course of dissipation.

The variation of the frequency shift and the wave amplitude with the normalized distance is shown in Fig.~\ref{Fig13}. 
As discussed above, the dissipation in the nonlinear regime should lead to stronger amplitude decay. 
Indeed, the nonlinear adiabatic evolution provides the most rapid amplitude decay. 
However, according to Fig.~\ref{Fig13}(a), in the case of a strong dissipation, the wavetrain amplitude decays slower with the distance than in the case of a weak dissipation. 
This apparent contradiction is explained by the scaling of the horizontal axis $\xi = \mu \omega_0^3 x$ that contains the dissipation coefficient $\mu$.
%***********************************

Though all the frequency shifts in Fig.~\ref{Fig13}(b) are not great in magnitudes, the eventual frequency shift within the linear theory shown with the green curve, is definitely larger in value than the simulated shifts within the nonlinear theory (blue curves). 
In the case of a weak dissipation, the simulated frequency shift (dashed blue curve) is close to the adiabatic solution (solid orange curve). 
The latter demonstrates the smallest downshift, while the simulation with strong dissipation (solid blue curve) is closer to the linear solution. 
However, the frequency evolution at the early stage looks more complicated (see the insertion in Fig.~\ref{Fig13}(b)). 
Initially, the frequency shift in the nonlinear adiabatic regime is more substantial than within the linear framework. 
%This conclusion holds for both the cases of weak and strong dissipation simulated numerically. 
The initial downshifting trend according to the adiabatic solution (\ref{Eq12_AlternativeForm}) (red dashed-dotted curve) agrees with the result of numerical simulation in the weak dissipation case only within a very short distance; but then, the nonlinear frequency downshifting quickly decelerates. 
The downshifting in the case of a strong dissipation (solid blue curve) is initially less rapid than in the adiabatic regime; it is initially closer to the linear solution but for some significant distance, its frequency is below the one in the linear framework. 
The approximate formula for the linear downshifting rate (\ref{PeakFrequencyVariation_Sech}) is found to describe the initial stage of the linear evolution very well, see the black dashed-dotted straight line which almost coincides with the green curve in Fig.~\ref{Fig13}(b).
Simulations of the NLS equation with the same dissipation term like in Ref.~\onlinecite{Alberello-22} give very similar results (see small empty circles in Fig.~\ref{Fig13}).

\section{DISCUSSION of results obtained}
\label{Sect-6}

Physical mechanisms of wave frequency downshifting when the water surface is covered by ice are discussed in this paper, related to the frequency-dependent dissipation coefficient. 
The linear mechanism has been discussed in Refs.~[\onlinecite{Keller-1998}] and [\onlinecite{Meylan-18}]; it consists of a faster decay of spectral areas with higher frequencies, therefore the downshift of the spectral peak is apparent; it is not related to the energy transfer between spectral intervals. 
Naturally, the frequency downshift in the linear scenario does not depend on the particular wave amplitude, but is controlled by the relative spectral width, see the estimates in Eqs.~(\ref{MeanFrequencyRate_NarrowBand}) and (\ref{PeakFrequencyVariation_Initial}). 
The broader the spectrum is, the more efficiently the spectral gradient of the dissipation rate produces the frequency downshifting. Note that the particular waveform (i.e. either the waves are grouped or spread in the physical domain) does not affect the spectral downshifting in the linear regime.

Redistribution of the spectral energy becomes possible when nonlinear wave-wave interactions are taken into account. Coherent long-lived wave groups can propagate in deep water, which in the leading order may be approximated by envelope solitons of the nonlinear Schr\"{o}dinger equation. We show that in the adiabatic nonlinear regime of transformation when the dissipation is weak compared to the nonlinearity, slowly decaying soliton groups experience downshifting. 
Similar to the linear framework, the downshifting effect takes place if the dissipation coefficient increases with frequency.

When the evolutions of a soliton-shape wavetrain in linear and nonlinear adiabatic regimes are compared, the frequency as a function of wavetrain amplitude decreases faster in the linear case (see Fig.~\ref{Fig12}), though the amplitude as a function of distance decreases faster in the nonlinear regime (see Fig.~\ref{Fig13}). 
The derivative of the local frequency to the distance is more substantial in the nonlinear adiabatic scenario (cf. Eq.~(\ref{PeakFrequencyVariation_Sech}) and Eq.~(\ref{Eq12_AlternativeForm})). 
In the course of the adiabatic evolution of a wavetrain, its characteristic sech-type shape is preserved in terms of surface displacement. 
Therefore, the estimates of the frequency downshift within the linear and nonlinear mechanisms (\ref{PeakFrequencyVariation_Sech}) and (\ref{Eq12_AlternativeForm}) hold for any instant of the adiabatically evolving envelope soliton.
However, solutions of the boundary problems (using the Fourier integral representation in the linear problem and using the numerical simulation in the nonlinear case) reveal the picture opposite to the analytic predictions: the eventual frequency shift in the linear framework is noticeably more significant than in the nonlinear adiabatic regime.
To obtain more representative estimates of the spectral evolution, we consider below the evolution of a relative spectral width.

\subsection{Evolution of the relative spectrum width of an envelope soliton} 
\label{sec:SpectrumWidthEvolution}

When an envelope soliton decays, its width increases; therefore, simultaneously with the frequency downshifting, the absolute spectral width of a soliton decreases, which prevents further downshifting.
The relative spectral width $\delta_m(\xi)$ which accounts for both these counteracting processes, estimated through the spectral moments and assuming that the distance $\xi$ is small, reads:
\begin{align} 
\label{SpectrumWidthEstimateAdiabatic}
\frac{\delta_m(\xi)}{\delta_m(0)} \approx \frac{A(\xi)}{A_0} \frac{\omega_0}{\omega_p(\xi)} \approx   1 - 2\left(1-n \delta_m^2(0) \right)  \xi .
\end{align}
Due to the adiabatic character of evolution, the approximation (\ref{SpectrumWidthEstimateAdiabatic}) reformulated for $\delta_m(\xi-\xi_0)$ is applicable at any instant of the evolution given $A_0 = A(\xi_0)$ and $\delta_m(0)=\delta_m(\xi_0)$ are the starting values of the amplitude and relative spectral width. 
According to Eq. (\ref{SpectrumWidthEstimateAdiabatic}), the relative spectral width of adiabatically decaying envelope soliton can formally grow when $n > \delta_m^{-2}(0)$. However, this condition requires a very high power $n$ of the dissipation law, even when the wave group is relatively short (for example, for the short group shown in Fig.~\ref{Fig03} it requires $n>48$). 
Hence, a realistic adiabatically evolving envelope soliton is characterized by a decaying relative spectrum width, and therefore the effect of downshifting diminishes with the distance in accord with the estimate (\ref{MeanFrequencyRate_NarrowBand}) which comes from the general formula (\ref{MeanFrequencyRate}) under the narrow-band assumption. This conclusion is in agreement with the adiabatic theory presented in Sec.~\ref{Sect-5}.

If the absolute spectral width is constant but the dominant frequency decreases, then the relative spectral width goes up and hence, the downshifting effect intensifies according to the formula (\ref{MeanFrequencyRate_NarrowBand}). 
Due to the frequency-dependent dissipation, the absolute spectrum width of waves evolving within the linear problem changes with distance as shown in Fig.~\ref{Fig03}. 
The estimate of the linear evolution of the relative spectral width of a sech-shape group (\ref{Eq26}) at the early stage can be obtained from Eq.(\ref{MeanFrequencyRate_NarrowBand}) by the direct calculation:
\begin{align} 
\label{SpectrumWidthEstimateLinear}
\frac{\delta_m(\xi)}{\delta_m(0)} \approx   1 - \frac{2n}{5} \left( 4n-9 \right) \delta_m^2(0) \xi .
\end{align}
Here it is assumed that $|\xi| \ll 1$ and the spectrum is relatively narrow, $\delta_m \ll 1$, where the mean spectral width of a sech-shape group at the initial moment is $\delta_m^2(0)=1/[3(\omega_0 T_0)^2]$. 
For the examples of short and broad wavetrains shown in Fig.~\ref{Fig03}, these values are $\delta_m(0) \approx 0.14$ and $\delta_m(0) \approx 0.04$ respectively. 
The estimate (\ref{SpectrumWidthEstimateLinear}) predicts that for $n \ge 9/4$ (what is true for the case of $n=3$) the relative spectral width decreases with the distance $\xi$. 
Note that this variation occurs very slowly due to the small parameter $\delta_m(0)$, in contrast to the adiabatic regime described by Eq. (\ref{SpectrumWidthEstimateAdiabatic}), where the coefficient at $\xi$ for realistic parameters is not small. 
Moreover, the direct calculation of $\delta_m(\xi)$ based on the linear solution (\ref{Eq26}) reveals that the small reduction of $\delta_m$ is shortly after changed by substantial growth of the spectral width.

\begin{figure}[h!]
\centerline{\includegraphics[width=0.8\textwidth]{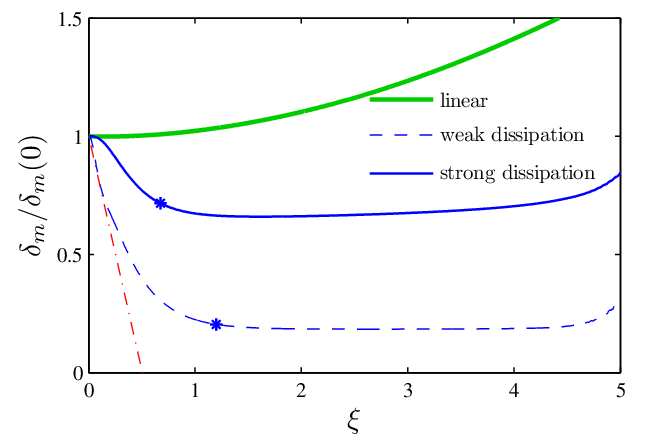}}
%\vspace{-0.5cm}
\caption{(Color online.) Relative change of the mean spectral width in the numerical simulations of the NLS equation (blue curves) compared to the linear simulation (green curve). The estimate for the early stage of an adiabatic evolution according to Eq.  (\ref{SpectrumWidthEstimateAdiabatic}) is given by the red dashed-dotted line.}
\label{fig:SpectrumWidthEvolution}
\end{figure}

In summary, one can conclude that in the course of the evolution of an envelope soliton, the relative spectral width decreases in the nonlinear adiabatic regime, but it increases in the linear setting. 
This explains the eventually faster downshifting within the linear problem observed in Fig.~\ref{Fig13}. 
The evolution of the relative spectral width is shown in Fig.~\ref{fig:SpectrumWidthEvolution} where the parameters $\delta_m$ are calculated based on the results of the numerical simulation of the NLS equation and the linear solution (\ref{Eq26}). 
The red dashed-dotted line represents the estimate for the adiabatic nonlinear regime (\ref{SpectrumWidthEstimateAdiabatic}) which agrees with the direct numerical simulation (dashed blue curve).
At small distances $\xi$ the curve for the strong dissipation case (solid blue curve) goes closer to the linear solution, which is consistent with the evolution of the peak frequency shown in Fig.~\ref{Fig13}(b)), but it decays at greater $\xi$ similar to the weakly dissipative case. 
At sufficiently large distances when due to the decay of the wave amplitude the dissipation becomes comparable with the nonlinearity (see the stars which correspond to the condition $A=A_l$), the functions $\delta_m(\xi)$ in the nonlinear problems start to grow. 
Remarkably, the corresponding intervals $A<A_l$ in Fig.~\ref{Fig12} (below the stars) are indeed characterized by faster downshifting. 
However, since the wave amplitude becomes very small at this stage, the late fast downshifting is unimportant in practice.

%***********************************

\subsection{Nonlinear spectral broadening}

From the above discussion, it becomes clear that a bigger frequency downshifting could occur in wavetrains of broader spectra. 
The adiabatic nonlinear evolution leads to spectral narrowing and thus, is less effective in downshifting compared to the linear regime when the relative spectrum width grows with time. 
The self-modulation of nonlinear waves is a recognized mechanism of wave spectrum broadening, see e.g. Ref.~\onlinecite{Chabchoubetal2013}. 
The self-modulation of weakly perturbed uniform wavetrains is examined most frequently in this respect, though for the present work, it is more convenient to consider the degenerate two-soliton solution of the NLS equation, which for the normalized equation (\ref{TNLS-Dimensionless}) with $\mu=0$ has the form \cite{Peregrine1983,Chabchoubetal2021}:
%
%\begin{align} \label{DoubleSolitonSolution}
%\Psi(\chi,\tau)=\Psi_D(\chi, \tau-\chi), \quad \Psi_D(\chi,\tau)=4 \frac{\tau \sinh{\tau}-\cosh{\tau}-2i\chi\cosh{\tau}}{\cosh{2\tau}+1+2\tau^2+8\chi^2}\exp{\left( i\chi \right)}.
%\end{align}
%
%
\begin{align} 
\label{DoubleSolitonSolution}
\Psi(\chi,\tau) = 4 \frac{\left(\tau-\chi\right) \sinh{\left(\tau-\chi\right)} - \left(1 + 2i\chi\right)\cosh{\left(\tau-\chi\right)}}{\cosh{2\left(\tau-\chi\right)} + 1 + 2\left(\tau-\chi\right)^2 + 8\chi^2}\exp{\left( i\chi \right)}.
\end{align}
This is an exact solution of the conservative NLS equation which reduces to two separated envelope solitons with identical carrier frequencies and amplitudes when $\chi \to \pm \infty$ (with the zero frequency and unit amplitude in terms of the solution given by Eq. (\ref{DoubleSolitonSolution})). 
The solitons interact and focus to a twice higher and shorter wave group when $\chi=0$ with the peak wave located at $\tau=0$. 
This solution was successfully reproduced in a laboratory tank and numerical simulation of primitive hydrodynamic equations in Ref.~\onlinecite{Chabchoubetal2021}. 
It was obtained that the wave evolution of moderate steepness is described by the solution (\ref{DoubleSolitonSolution}) reasonably well.

\begin{figure}[h!]
\centerline{\includegraphics[width=0.8\textwidth]{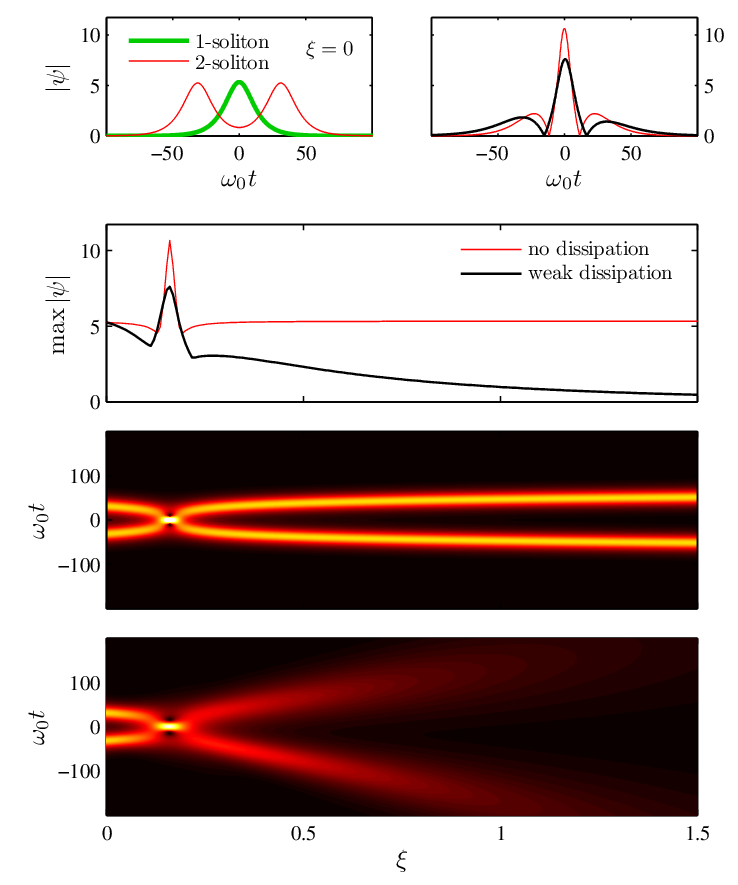}}
%\vspace{-0.5cm}
\caption{(Color online.) A comparison of the exact degenerate two-soliton solution with the numerical solution of the NLS equation with a weak dissipation. Top row, left: the initial condition (thin red curve) compared to the single soliton solution (thick green curve). 
Top row, right: wave envelopes when the maximum amplitudes are achieved according to the exact solution (\ref{DoubleSolitonSolution}) (thin red curve) and in the simulation (thick black curve). 
Second row: evolution of the envelope maxima according to the exact solution (thin red curve) and in the simulation (thick black curve). 
Third and fourth rows: space-time diagrams of $|\psi(\xi,t)|$ according to the exact solution and numerical simulation, respectively.}
\label{fig:DoubleSoliton}
\end{figure}

We performed numerical simulations of the NLS equation with a weak dissipation, $\mu = 0.02$~s$^{-1}$ and other parameters as per Eq.~(\ref{param}). 
For the initial condition, we took the degenerate solution (\ref{DoubleSolitonSolution}) at the moment preceding the focusing when it describes two weakly overlapping envelope solitons with similar amplitudes $A_0$ and carrier frequencies $\omega_0$, see the thin red curve in the left-top panel of Fig.~\ref{fig:DoubleSoliton}. 
In that figure for comparison, a single soliton is shown by the thick green curve. 
Thus, the only noticeable difference of the initial condition from a single envelope soliton, is that now there are two almost independent solitons. 
With this initial condition, the analytic solution predicts a focusing at the distance $x_f=2000$ m ($\xi_f\approx 0.16$). 
The focusing is reproduced in the weakly dissipative case too, but with a smaller value of the wave amplitude, see the dependences of wave maxima on the distance $\xi$ in the second frame of Fig.~\ref{fig:DoubleSoliton}; the envelope shapes at the focusing points are shown in the top-right panel of the figure.
The general views of the evolution according to the exact solution and in the numerical simulation of the dissipative NLS equation are shown in space-time pseudocolor diagrams in the two lowest panels of Fig.~\ref{fig:DoubleSoliton}. 
The soliton focusing corresponds to the bright spots at $\xi\approx 0.16$.

%%%%%%%%%%%%%%%%%%%%%%%%%%%%%%%%%%%%%%%%%%%%%%%%%%%%%%%%
\begin{figure}[b!]
\vspace{-0.9cm}
\centerline{\includegraphics[width=0.8\textwidth]{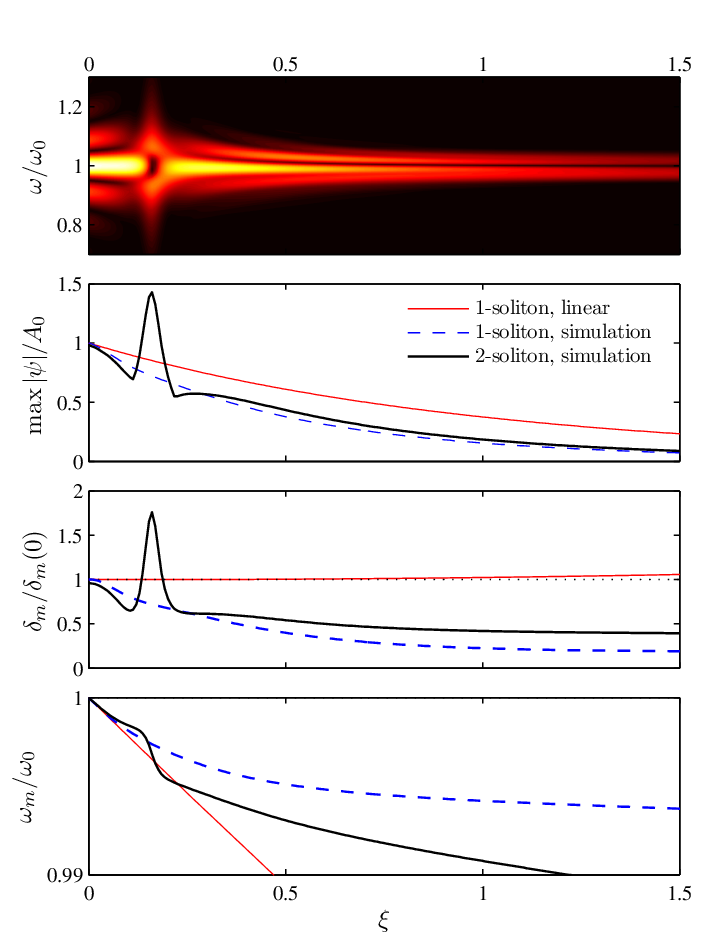}}
\vspace{-0.5cm}
\caption{(Color online.) Weakly dissipative evolution of two interacting solitons shown in  Fig.~\ref{fig:DoubleSoliton}. 
Top: evolution of the temporal Fourier spectrum. 
The remaining three panels demonstrate, respectively, the evolution of the maximum wave amplitude, relative spectral width, and mean frequency, compared to the cases of a single soliton within the linear and nonlinear frameworks, see the legends.}
\label{fig:DoubleSolitonSpectrum}
\end{figure}

The spectral domain of the simulated evolution of the two-soliton wavetrain is examined in Fig.~\ref{fig:DoubleSolitonSpectrum}. 
Amplitudes of the temporal Fourier transform as functions of the distance are shown in the upper panel of the figure. 
Initially, in the Fourier transform there is one dominant peak at $\omega = \omega_0$; and the spectrum is close to the one of a single envelope soliton. 
The Fourier spectrum broadens near the focusing point and for some short time becomes two-humped. 
After that, the spectrum shrinks back but its single peak splits into two peaks at a large distance. 
These two peaks correspond to two separated wave groups (see in the bottom panel in Fig.~\ref{fig:DoubleSoliton}); these wave groups have slightly different carrier frequencies and amplitudes.

Variation of the envelope maximum with the distance is presented in the second panel of Fig.~\ref{fig:DoubleSolitonSpectrum}. 
In the same panel, we show spatial variation of the single soliton amplitude as per numerical simulation and a linear wavetrain (according to Eq.~(\ref{Eq26})). 
These three cases were also examined for the relative spectral width $\delta_m$ and the mean frequency $\omega_m$; the results are shown in the two lowest panels of Fig.~\ref{fig:DoubleSolitonSpectrum}. 
Due to the splitting of dominant wave frequency in the course of the wave evolution, here we use mean values of the frequency and frequency width as defined through the spectral moments in Sec.~\ref{sec:GeneralLinearTheory}.
Note that in the figure, for all three cases, the widths $\delta_m(\xi)$ are normalized with the same initial spectral width $\delta_m(0)$ calculated for the initial condition in the form of a single soliton; the amplitude parameter $A_0$ in case of the two-soliton solution corresponds to the amplitude of solitons when they are very far from each other. 
One can see from Fig.~\ref{fig:DoubleSolitonSpectrum} that at locations far from the place of focusing, the evolution of the wave maximum in the case of interacting solitons is generally close to the case of a single soliton. 
Recall that the latter may be well approximated by the adiabatic theory since the dissipation coefficient is small. 
Initially, the relative spectral width in the case of two solitons is a bit smaller than that of a single soliton. 
The early stage of evolution of interacting solitons proceeds according to the nonlinear adiabatic scenario; the spectral width decreases. 
However, the focusing process is accompanied by a quick and strong broadening of the spectrum; then, after the focusing, the spectrum narrows back and continues to narrow further similar to what happens in the nonlinear adiabatic case. 
Thus, one can say that after the collision, solitons continue to evolve adiabatically. 
The spectrum width in the linear framework becomes a little broader at $\xi=1.5$ in agreement with the discussion in the previous subsection \ref{sec:SpectrumWidthEvolution}.

In accord with the discussion in Sec.~\ref{sec:SpectrumWidthEvolution}, the variation of the relative spectrum width determines the variation of the dominant frequency, cf. the two bottom panels in Fig.~\ref{fig:DoubleSolitonSpectrum}. 
Though in general, the dominant frequency of interacting solitons decreases slower than in the linear case, during the nonlinear collision, the frequency experiences a fast substantial downshift, so that for some interval of distances the dominant frequency of interacting solitons becomes even smaller than in the linear regime.

Since the picture of spectral transformation is found to be rather complicated, and the values of peak and mean frequencies, and the spectral width estimates may appear particular, we have also analysed the integrated characteristic of the energy flux in the Fourier space in the Appendix. As was aforementioned, in the linear regime, there is no energy exchange between spectral components in the course of wave propagation and dissipation, therefore the energy flux is in fact apparent and illusive. This is obviously not the case in the nonlinear regime. According to the analysis presented in the Appendix, the character of energy fluxes in different regimes of wave evolution is consistent with the already drawn conclusions: the flux in the linear regime is more substantial compared to the nonlinear quasi-adiabatic scenarios, though the nonlinear spectral spreading caused by interactions of coherent wave groups can change the situation to the opposite.

\section{CONCLUSION}
\label{Sect-7}

In this paper, we have presented a physical explanation and calculation of the frequency downshift within wavetrains in ice-covered seas. 
The downshifting is caused by the frequency-dependent dissipation with larger decaying rates in higher frequencies.
A detailed examination of the frequency downshifting has been carried out within the linear and nonlinear models for the typical sech-shaped wavetrain.
As the convenient mathematical model, the dissipative nonlinear Schr\"odinger equation for weakly nonlinear narrow-band spectrum waves was exploited.
Analytical estimates of the downshift rates have been obtained within the approximated adiabatic theory when the dissipation is weak. 
The long-term wave evolution was numerically simulated to verify the analytic results and to consider the regime when the nonlinear wave evolution is not adiabatic.

For the typical physical parameters of weakly dissipating due to the ice oceanic waves (\ref{param}) adopted from Ref.~[\onlinecite{Alberello-22}] the characteristic spatial scales of the wave attenuation (defined as $\xi=1$) are of the order of 120~km and 2.2~km for the frequencies $\omega_0=2\pi/12$~rad/s and $\omega_0=2$~rad/s respectively, for the power $n=3$. In the ``strong dissipation case'' these distances are ten times shorter.

It has been shown that the eventual frequency shift and the corresponding apparent spectral energy flux are noticeably more significant in the linear framework than in the nonlinear adiabatic regime for envelope solitons. 
The nonlinear evolution within the non-adiabatic regime of wave evolution (for example, when the dissipation is strong) exhibits an intermediate result with the frequency downshift larger than in the adiabatic nonlinear regime but still smaller than in the linear case. 
Note that the degree of nonlinearity is determined by the ratio between the modulating effect of the nonlinearity in deep water and the counteracting dispersion which may be quantified by the Benjamin--Feir Index \cite{Onoratoetal2001}.

However, the spectral downshifting is much accelerated and may exceed the downshifting rate in the linear regime, when envelope solitons collide and create temporal wave groups which are higher in amplitude and, most importantly, broader in the spectrum. 
The degenerate two-soliton solution of the NLS equation was considered in this work just for the sake of convenience since at the asymptotic stage of the evolution far from the place of their collision, the solution represents just a pair of identical envelope solitons. 
It is important to note that the considered example is just a particular case of a more general solution of the conservative NLS equation called bi-soliton \cite{Peregrine1983} (or bound state) which describes a repeating beat between two envelope solitons with the same carrier frequency but different amplitudes. 
Such beatings are accompanied by a substantial broadening of the spectrum. 
In a slightly more generalized setting when two envelope solitons have slightly different amplitudes and frequencies, the process of collision looks very similar. 
In the meantime, the degenerate two-soliton solution can be considered as a counterpart of a breather solution of the NLS equation (see Ref.~\onlinecite{Chabchoubetal2021}), which is a recognized source of strong spectral broadening and is a consequence of the modulational instability \cite{Dudleyetal2009,Chabchoubetal2013}. 
The spectral up- and down-shifting of NLS breather solutions affected by wind were observed in the laboratory and simulated numerically in Ref. \onlinecite{Eeltinketal2017}, where a higher-order NLS equation was modeled with frequency-dependent terms responsible for the viscosity and wind.
Therefore, we anticipate that spectral downshifting should be strongly intensified when long-lived coherent nonlinear wave groups collide or interact with any waves of significant amplitude. 
This can occur under the conditions of emerging modulational (Benjamin -- Feir) instability \cite{ShemerSergeeva2009,SlunyaevSergeeva2011} or when the soliton-like nonlinear groups (representing the so-called soliton gas \cite{Suretetal2020}) are characterized by different carrier frequencies and thus collide due to the difference in velocities.

In conclusion, we emphasize that the particular dependence of the dissipation coefficient on the frequency $\omega^n$ modifies the process, but all the qualitative outcomes hold regardless of the power $n$. 
Therefore the results obtained are not restricted to the situation of an ocean covered by ice floes but have much broader application.

\begin{acknowledgments}
Secs.~\ref{Sect-5}, \ref{Sect-6} are performed with the support of the Russian Science Foundation, grant No 22–17-00153. 
The research reported in the remaining sections was supported by the Ministry of Human Resource and Development, Government of India through Apex Committee of SPARC vide Grant No. SPARC/2018-2019/P751/SL, and also by the Ministry of Science and Higher Education of the Russian Federation (Grant No. FSWE-2023-0004); and by the Council of the grants of the President of the Russian Federation for the state support of Leading Scientific Schools of the Russian Federation (Grant No. NSH-70.2022.1.5). \\
\end{acknowledgments} 

\noindent {\bf AUTHOR DECLARATIONS} \\

\noindent {\bf Conflict of Interests:} The authors have no conflicts to disclose. \\

\noindent {\bf Author Contributions:} \\
{\bf Alexey Slunyaev}: Conceptualization (equal); Formal analysis (equal); Investigation (equal); Methodology (equal); Software; Validation (equal); Visualization (equal); Writing -- original draft (equal); Writing -- review \& editing (equal). {\bf Yury Stepanyants}: Conceptualization (equal); Formal analysis (equal); Investigation (equal); Methodology (equal); Validation (equal); Writing -- original draft (equal); Writing -- review \& editing (equal).\\

\noindent {\bf Data Availability:} The data that support the findings of this study are available from A.S. upon a reasonable request.

\appendix
\section{SPECTRAL ENERGY FLUX}
\label{Appendix}

Given the intricate picture of the spectral evolution of wave groups under the effect of frequency-dependent dissipation, let us estimate the related energy flux in the Fourier domain. 
This quantity can be defined in a normalized form as follows,
\begin{align} 
\label{EnergyFlux}
f(\xi) = \frac{1}{\omega_0}\frac{J_0(\xi)}{J_0(0)} \frac{d \omega_m}{d\xi},
\end{align}
where $J_0$ is the zeroth spectral moment defined in Eq. (\ref{SpecMoments}).
Owing the analytic expression for the evolution of $\omega_p(\xi)$ rather than $\omega_m(\xi)$, its is constructive to use an alternative definition $f_p$:
\begin{align} 
\label{EnergyFluxViaPeak}
f_p(\xi) = \frac{1}{\omega_0}\frac{J_0(\xi)}{J_0(0)} \frac{d \omega_p}{d\xi}
\end{align}
as the rough estimation of the exact energy flux, $f(\xi) \approx f_p(\xi)$.

\begin{figure}[b!]
\centerline{\includegraphics[width=1.0\textwidth]{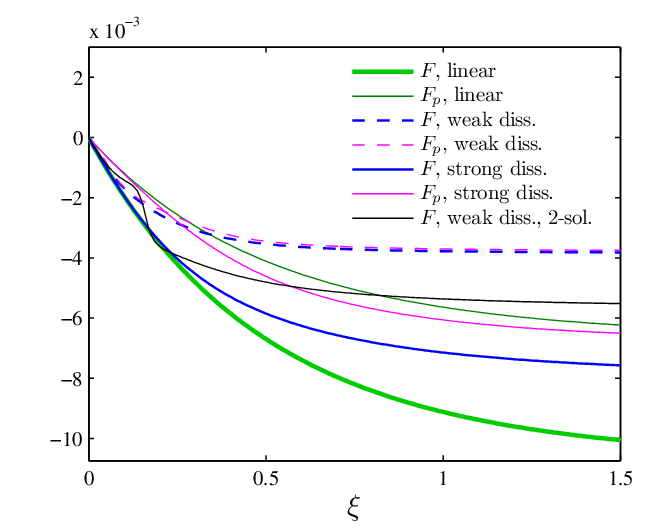}}
\vspace{-0.5cm}
\caption{(Color online.) Accumulated over a distance energy fluxes $F$ and their approximate estimates using the peak frequency $F_p$ within the analytic linear theory and in the numerical simulations of the NLS equation with the frequency-dependent dissipation.}
\label{fig:Fluxes}
\end{figure}
%***********************************

Numerically computed integrals of the spectral energy fluxes over the normalized distance, $F(\xi)=\int\limits_0^{\xi}{f(\chi) d\chi}$ and  $F_p(\xi) = \int\limits_0^{\xi}{f_p(\chi) d\chi}$, are shown in Fig.~\ref{fig:Fluxes}. 
The linear solution for the wave group of the shape of an envelope soliton (\ref{Eq26}) exhibits remarkable difference between the exact solution $F$ (thick green curve) according to Eq.~(\ref{EnergyFlux}) and the estimation $F_p$ using the peak frequency Eq.~(\ref{EnergyFluxViaPeak}) (thin green curve), what is in qualitative agreement with the different behavior of $\tilde{\omega}_m$ and $\tilde{\omega}_p$ shown in Fig.~\ref{Fig04}. 
At a large distance, this leads to about two times the underestimation of the energy flux when the peak frequency is used.

In the nonlinear adiabatic regime with small dissipation, the fluxes $F$ and $F_p$ practically coincide (see the blue and magenta dashed lines). 
This is a consequence of the preservation of the symmetric shape of a wavetrain in the course of its propagation and dissipation. 
Albeit the energy flux in the nonlinear adiabatic regime initially decays stronger than in the linear regime with the approximation $F_p$ (sf. blue and magenta dashed lines with the thin green line; see also the green line in Fig.~\ref{Fig04} for the peak frequency), it slows down at larger distances and eventually becomes about trice less than the flux $F$ in the ``exact'' linear case (see the thick green line).

In the nonlinear simulation with strong dissipation, the energy flux calculated with the approximate formula (\ref{EnergyFluxViaPeak}) using the peak frequency $F_p$ significantly underestimates the real flux (cf. solid magenta line and solid blue line). 
At the initial stage for some distance, the nonlinear energy flux $F$ (solid blue line) coincides with the flux in the linear regime (thick green curve), though later it reduces and finally demonstrates a significantly smaller absolute value compared to the linear case. 
Overall, the linear flux (thick green curve) corresponds to the most significant energy redistribution towards low frequencies.

The flux calculated in the simulation of two interacting envelope solitons is shown in Fig.~\ref{fig:Fluxes} by the solid black line. 
Its dependence on distance is in agreement with the behavior of related frequency widths presented in Fig.~\ref{fig:DoubleSolitonSpectrum}. 
Albeit the flux is small when the solitons are distant, the interaction between them leads to the formation of a larger and shorter wavetrain which yields a significantly faster transfer of energy to the low-frequency range, which can be even more significant than in the linear case.

%\bibliography{MSS}
% 

\end{document}